\documentstyle{article}
\begin{document}
\baselineskip=24truebp
\newcount\secnum
\newcount\eqnum
\newcount\subnum
\eqnum=0\secnum=0\subnum=0
\def\eqnoi{\global\advance\eqnum by 1}
\def\chapitre#1{\global\advance\secnum by
1\eqnum=0\subnum=0\bigskip\bigskip\centerline{\bf{\the\secnum}:
#1}\bigskip\medskip\noindent}
\def\subsec#1{\global\advance\subnum by 1\medskip
{\bf\centerline{\the\secnum.\the\subnum: #1}\medskip\noindent}}
\def\back#1{{\advance\eqnum by-#1(\the\secnum.\the\eqnum)}}
\def\last{(\the\secnum.\the\eqnum)}
\def\appendice#1{\eqnum=0\subnum=0\bigskip\bigskip\def\appnum{\char`#1}
\centerline
{\bf{Appendix #1}}
\bigskip\medskip\noindent}
\def\eqnoa{\global\advance\eqnum by 1}
\def\backa#1{{\advance\eqnum by-#1(\appnum.\the\eqnum)}}
\def\lasta{(\appnum.\the\eqnum)}
\def\otpi{{1\over2\pi}}
\def\tr{\hbox{\rm Tr}\,}
\def\pr#1{#1^\prime}
\def\ppr#1{#1^{\prime\prime}}
\def\pppr#1{#1^{\prime\prime\prime}}
\def\tpipq{{2\pi ip\over q}}
\def\tpips{{2\pi ip\over s}}
\def\tpiq{{2\pi i\over q}}
\def\ccc{\hbox{\rm c. c.}}
\def\eps{\epsilon}
\def\avg#1{\left\langle#1\right\rangle}
\def\part#1#2{{\partial#1\over\partial#2}}
\def\jac#1#2{{\left(#1\over#2\right)}}
\def\mq{\left|M_q\right|}
\def\ms{\left|M_s\right|}
\def\mqr{\left|M_{q,r}\right|}
\renewcommand{\theequation}{\the\secnum.\the\eqnum}
\begin{titlepage}
\title{{\bf Distribution of Eigenvalues for the Modular Group}}
\author{E. Bogomolny
\thanks{On leave of ansence from L.D. Landau Institute of Theoretical Physics,
142432 Cherogolovka, Russia}
\and F. Leyvraz
\thanks{Permanent address: Instituto de F{\'\i}sica, University of Mexico,
Apdo.~postal 20--364, 01000 Mexico City, Mexico} \and C. Schmit}
\date{Division de Physique Th\'eorique\thanks{Unit\'e de Recherche
des Universit\'es Paris 11 et
Paris 6, Associ\'ee au CNRS}\\
Institut de Physique Nucl\'eaire\\
91406 Orsay Cedex, France}
\maketitle
\thispagestyle{empty}
\begin{abstract}
The two-point correlation functions of energy levels for
free motion on the modular domain, both with periodic
and Dirichlet boundary conditions, are explicitly computed
using a generalization of the Hardy--Littlewood method.
It is shown that in the limit of small separations they show
an uncorrelated behaviour and agree with the Poisson
distribution but they have prominent
number-theoretical oscillations at larger scale. The results
agree well with numerical simulations.
\end{abstract}

\vspace{2cm}

{\bf IPNO/TH 94--43}\begin{flushright} {\bf June 1994}\end{flushright}

\end{titlepage}
\chapitre{Introduction}
Free motion on constant negative curvature surfaces (CNCS)
generated by discrete groups is the oldest and in some sense the best
example of classically chaotic motion (see e.g.
\cite{{hopf},{enc},{balvor}}).
In recent years this subject has attracted wide attention also within
the context of quantum chaos. There the main question is the way in
which classical chaos manifests itself in the properties of the
corresponding quantum systems (see e.g. \cite{{houches},{gutzbook}}).
An important property of CNCS models is the existence of an
{\em exact} relationship between the density of eigenvalues of
the Laplace--Beltrami operator on the surface (= energy levels)
and the geodesic
on the surface, which correspond to classical periodic orbits.
This is known as the Selberg trace formula (see e.g.
\cite{{selberg},{hejhal}}). For arbitrary systems, only an
approximate connection of this type is known, namely the so-called
Gutzwiller trace formula \cite{{gutz},{gutzbook}}, which is asymptotically
valid in the limit of highly excited states but does not have a good estimate
on the error. It is therefore the coexistence of hard classical chaos
and the exact Selberg trace formula which makes the study of CNCS models
so important.

The simplest hallmark of classical chaos for a quantum system is
the nature of the spectral fluctuations of the energy levels.
It was conjectured that, for ergodic systems with strong
chaotic properties, the fluctuation properties
of energy levels should be
that of the classical random matrix ensembles \cite{bgs}.
This result has found considerable numerical confirmation, but
numerical work on various models on CNCS
\cite{{charles},{charles1},{aurich1},{aurich2}}
gave unexpected results. It was observed that the distribution of
energy levels for these systems is quite close to a Poisson
distribution, normally typical of integrable systems \cite{berrytab}
and not to random matrix ensembles which are characterized
by strong level repulsion.

In \cite{{georgeot},{steiner}}, it was shown that this anomalous
behaviour could be traced back to a non-generic feature of these
systems, namely the fact that the underlying group belonged to
a very specific subclass of the discrete subgroups of the motions
of CNCS, namely the so-called arithmetic groups.

Arithmetic groups are groups which permit a representation by
$n\times n$ matrices with integer entries \cite{gelfand}.
The important consequence of the arithmetic nature
of these groups is that in such cases the corresponding CNCS
shows an exponential proliferation of geodesics having
exactly degenerate lengths \cite{{georgeot},{georgeot2}}.
It is the cumulative effect of the interference of these
degenerate orbits which leads to the Poisson-like distribution
of energy levels. In \cite{{georgeot},{georgeot2}}
the two-point correlation function of energy levels
was computed in the diagonal approximation \cite{berry},
where one neglects all correlations between orbits that are
not exactly degenerate. It was shown that this function definitely
differs from the result predicted by random matrix theory
\cite{mehta}. Unfortunately, the diagonal assumption
is quite crude and is only expected to give good results when the separation
between two energy levels entered two--point function is large.
In particular, it does not allow one
to compute the correlations in the region where these are believed
to be universal, namely for energy differences of the order
of a level spacing.

The purpose of this paper is to compute explicitly the two-point correlation
function for the energy levels of the modular domain and the corresponding
billiard. The calculations are based on a generalization of the
Hardy--Littlewood  method \cite{hardy} which was originally developed to
compute the distribution of prime numbers,
and depend strongly
on the number-theoretical properties of the multiplicities of the
periodic orbits of the modular group.

This paper is organized as follows: In Sec.~2, we give a quick
review of hyperbolic geometry and the
Selberg trace formula and derive various
basic relations between the spectrum (in particular its
two-point function, which is the basic object of interest
here) and the properties of the classical periodic orbits
of the system. In Sec.~3, we describe a method originally due to
Hardy and Littlewood to describe the structure of singularities
of certain peculiar power series. This leads to the consideration
of certain quantities which are evaluated exactly. These results
are presented in detail in Sec.~4. This leads to a closed form for
the two-point form factor of the spectrum of the modular domain
which has $\delta$--function singularities at all rational points.
 In Sec.~5, we demonstrate that after a suitable smoothing it becomes
constant as it should for the Poisson distribution. This result is
valid when the separation is fixed and energy tends to infinity. On a scale
of $\ln k/k$ (where $k$ is the momentum)
the two-point form factor has oscillations
of number-theoretical origin. In Sec.~6 we generalize these formulas for the
case of modular billiard with Dirichlet and Neumann boundary conditions. In the
conclusion (Sec.~7) we briefly repeat the
main steps necessary for our derivation.
In Appendices~A--F the details of calculations are presented. Much
standard material in hyperbolic geometry and number theory has been presented
to make the article self-contained.

\chapitre{Basic Identities}
\setcounter{equation}{0}

In what follows we shall use the standard realization of the
surface of constant negative curvature as the so-called Poincar\'e
plane, that is, the upper half-plane $z=x+iy$, where $y>0$,
endowed with the metric
$ds^2=y^{-2}(dx^2+dy^2)$, (see e.g. \cite{{balvor},{magnus},{harm}}). The
hyperbolic distance between two points $z_1=x_1+iy_1$ and $z_2=
x_2+iy_2$ is given by
\begin{equation}
\eqnoi
\cosh d(z_1,z_2)=1+{|z_1-z_2|^2\over2y_1y_2}
\end{equation}
The geodesics are then circles perpendicular to the real axis
and the group of isometries is the group of linear fractional
transformations, that is
\begin{equation}
\eqnoi
\pr{z}=gz={az+b\over cz+d}
\end{equation}
where $a$, $b$, $c$ and $d$ are arbitrary real numbers, which can,
without loss of generality, be chosen so as to satisfy the
condition $ad-bc=1$. It is then easily verified that the
composition of two such transformations gives another such
transformation according to the multiplication of the corresponding
matrices
\begin{equation}
\left(\matrix{a&b\cr
c&d}\right)
\eqnoi
\end{equation}
One can now introduce finite surfaces through a device which has
an analogue in the Euclidean case: There, one can construct a torus
by identifying all points on the plane which differ from
each other by a translation belonging to a discrete subgroup of
the Euclidean group (which is the isometry group of the Euclidean
plane). In the case of CNCS
one proceeds similarly: One takes a given discrete subgroup $G$ of
$SL(2,R)$ and identifies all points which are connected by a transformation
$g$ belonging to $G$.

Under certain  assumptions, this procedure leads to a CNCS
on which free motion is ergodic. This motion has very strong
chaotic properties \cite{hopf,balvor,enc} and is therefore a natural
object of study.

To visualize this construction, it is convenient to introduce the
notion of the fundamental domain
\def\fd{fundamental domain}
of a given discrete group $G$. This is defined as a region in the upper
half-plane such that
\begin{enumerate}
\item No two points inside the \fd{} are connected by a transformation
$g$ belonging to $G$.
\item For any point $\pr{z}$ outside of the \fd, there is a point
$z$ inside the \fd{} such that there is a $g$ in $G$ with
\begin{equation}
\pr{z}=gz.\eqnoi
\end{equation}
\end{enumerate}
Eq.~\last{} leads to the identification of certain points on
the {\em boundary} of the \fd. Gluing these together yields
a compact surface of constant negative curvature. The trajectory on
the whole upper half-plane is given by a half-circle perpendicular
to the real axis. After the identification has been carried out,
this geodesic can be reduced to a curve lying entirely within the
\fd{} and consisting of segments of geodesics.

Periodic orbits of this geodesic flow are in one-to-one
correspondence with conjugacy classes of elements of $G$
(see e.g. \cite{balvor}). If $M$ is any matrix belonging
to $G$, then the (hyperbolic) length of the corresponding
periodic orbit is given by the relation
\begin{equation}
2\cosh l/2=|\tr M|
\eqnoi
\label{2.5}
\end{equation}

The natural ``quantization'' of such systems consists in the
investigation of the spectrum of the invariant Laplace--Beltrami
operator
\begin{equation}
-{y^2\over2}\left(\part{{}^2}{x^2}+\part{{}^2}{y^2}\right)\Psi_n(x,y)=
E_n\Psi_n(x,y)
\label{laplace}
\eqnoi
\end{equation}
on the space of functions obeying the periodic boundary conditions
\begin{equation}
\eqnoi
\label{bc}
\Psi_n(gz)=\Psi_n(z)
\end{equation}
for any element $g$ of $G$. (Note that our definition of $E_n$
differs by a factor of one half from the one commonly used
in the literature.) Due to their peculiar mathematical
structure, there exists for these models an exact relation between
``quantum'' eigenvalues and the periodic orbits of the corresponding
classical motion. This relation---the celebrated Selberg trace formula
\cite{{selberg},{hejhal}}---can be stated in the following way: Let
$h(r)$ be an arbitrary even
function with appropriate smoothness properties
and $g(u)$ its Fourier transform, given by
\begin{equation}
g(u)=\otpi\int_{-\infty}^\infty h(r)e^{-iur}dr.\eqnoi\end{equation}
Let the index $p$ run over all classical periodic orbits, let
$l_p$ denote the length of the corresponding orbit and $L_p$
be the length of the {\it primitive\/} periodic orbit.
Let $A$ be the area of the
fundamental domain under consideration. One then has the identity:
\begin{eqnarray}
\sum_{n=1}^\infty h(r_n)&=&{A\over4\pi}\int_{-\infty}^\infty
rh(r)\tanh\pi r\,dr+\sum_p {L_p\over2\sinh l_p/2}g(l_p)+\nonumber\\
&&\qquad+\hbox{
``corner and horn'' terms},\eqnoi
\end{eqnarray}
where $r_n$ is equal to $\sqrt{2E_n-1/4}$, with $E_n$ being the
eigenvalues of the Laplace--Beltrami
operator with periodic boundary conditions.

The first terms correspond to the smooth part of the level density
and the second gives the contribution from periodic orbits. In mathematical
language, it equals the sum over all conjugacy classes of matrices
with trace larger than two.
The ``corner and horn'' terms refer to contributions from group matrices
with trace less than and equal to two respectively, whenever such
elements exist.
Their explicit form can be found in \cite{{selberg},{hejhal}}.

In the following, we shall apply the Selberg trace formula to
\begin{equation}
h(r)=\delta(E-{1\over2}(r^2+1/4)).\eqnoi\end{equation}
This formally causes some problems, as the delta functions
are not smooth enough for the series to converge. We will here
proceed in a largely formal manner and will discuss the regularization later.

In this way one obtains for the eigenvalue density $d(E)=\sum_{n=1}
^\infty\delta(E-E_n)$
\begin{equation}
d(E)=\avg{d(E)}+d_{osc}(E)+\tilde d(E),
\eqnoi\end{equation}
where the first term is given $A/(2\pi)$
\begin{equation}
d_{osc}(E)=\sum_p {l_p\over\pi k}\sum_{n=1}^\infty{\cos kl_pn
\over2\sinh l_pn/2}.\eqnoi\label{basic}
\label{2.12}
\end{equation}
and $\tilde d(E)$ includes all other terms which enters the exact Selberg trace
formula.
Here $k$ is the momentum, defined by $E=k^2/2+1/8$.

Without the last term the Selberg trace formula agrees with the
Gutzwiller trace formula. In principle, one could use a suitable
regularization of the delta functions and take the appropriate limits
at the end of the calculations.

So far, the above formulae are valid for an arbitrary group.
Now let us consider the simplest example of an arithmetic
group, namely the
modular group. This group is defined as the group
of all $2\times2$ matrices
with integer entries and unit determinant modulo the subgroup
$\{{\bf1},{\bf-1}\}$. It is well-known (see e.g. \cite{harm})
that this group is
generated by two of its elements: The translation
$T$, which maps $z$ to $z+1$ and the inversion $S$, which maps
$z$ to $-1/z$, with their corresponding matrices:
\begin{equation}
s=\left(\matrix{1&1\cr0&1\cr}\right)\qquad
t=\left(\matrix{0&1\cr-1&0\cr}\right).\eqnoi
\end{equation}
The \fd{} of the modular group has the form shown in Fig.~1 and the
area $\pi/3$. The arrows indicate the lines identified under the
action of the generators $T$ and $S$ and the
periodic boundary conditions of Eq.~(\ref{bc}) mean that the
function $\Psi(x,y)$ takes the same values on corresponding lines.
For this problem there is an evident  reflection symmetry
$x\to-x$, which leads to a splitting
of the eigenfunctions in two classes, namely odd and even. These
two classes correspond to the problem of finding the eigenvalues
of the Laplace--Beltrami operator with Neumann and Dirichlet
conditions on the boundary of the \fd{} shown in Fig.~1.
This problem is called the modular billiard (or Artin billiard
\cite{matth}) and we shall consider it in Sec.~6.

We have said that the modular group is the group of all $2\times2$
matrices with integer entries and unit determinant. The periodic
orbits on the modular domain then correspond in a unique way to
the conjugacy classes of those elements of the modular group which
have trace larger than two (hyperbolic elements). Further one
has the general relation
\begin{equation}
|\tr M|=2\cosh l_p/2.\eqnoi
\end{equation}
which connects the length of the periodic orbit $l_p$ with the
trace of a representative matrix of the conjugacy class. But as all
entries of the matrix are integers, the trace is also an integer.
Here the arithmetical nature of the group comes into play.

Therefore one sees that for all periodic orbits with a length
less than $L$, the number of possible different lengths is the number
of different integers $N$ less than $2\cosh L/2$, or asymptotically
as $L\to \infty$, $N$ goes as $e^{L/2}$.
Now it is well known that for any group the number of
periodic orbits of length less than $L$ grows as
\cite{{selberg},{hejhal},{huber}}
\begin{equation}
\eqnoi
N(l_p<L)=\frac{e^L}{L}
\end{equation}
up to exponentially smaller terms.
These two estimates show that in the case of the modular group
(as well as other arithmetic groups) periodic orbits are degenerate,
i.e., there are many periodic orbits with exactly the same length.

Let $g(l)$ be the number of periodic orbits of length $l$. The
above estimates mean that asymptotically
\begin{eqnarray}
\sum_{l<L}g(l)&=&{e^L\over L}\nonumber\\
\sum_{l<L}1&=&e^{L/2},
\eqnoi
\end{eqnarray}
where the summation extends over different lengths of periodic orbits, counted
without taking multiplicity into account. If we now define
the {\em mean} multiplicity $\avg{g(l)}$ as follows
\begin{equation}
\eqnoi
\avg{g(l)}={\hbox{Number of periodic orbits with $l<l_p<l+\Delta l$}
\over\hbox{Number of different lengths with $l<l_p<l+\Delta l$}},
\end{equation}
one concludes that
\begin{equation}
\avg{g(l)}=2{e^{l/2}\over l}.
\eqnoi
\end{equation}
For other arithmetic groups one obtains the same asymptotic behaviour
but with a numerical prefactor which depends on the group
\cite{{georgeot},{georgeot2},{bolte}}.
This extraordinary degeneracy of the
lengths of periodic orbits has been discussed before and is at the root
of the remarkable structure to be found in these systems.

We now proceed to reexpress Eq.~(\ref{basic}) in terms of a sum over
all conjugacy classes of hyperbolic elements of the modular group.
Denote by $n$ the trace of a given conjugacy class and by $g(n)$
the number of distinct conjugacy classes corresponding to trace $n$.
Taking into account the fact that $n$ goes as $e^{L/2}$ as $n\to\infty$
one concludes that
\begin{equation}
\avg{g(n)}={2e^{L/2}\over L}={n\over\ln n}.
\eqnoi
\end{equation}
Therefore the mean multiplicity of periodic orbits having trace $n$
grows asymptotically as the number of primes less than $n$.
While such a fact is suggestive of a deeper connection, the authors are
as yet unable to state anything further. The multiplicity $g(n)$ can
also be identified with the proper class number of quadratic
forms (\cite{harm,sarnak}), but we shall not require this
representation in the following.

Applying the Selberg trace formula to $d(E)$, we split the contributions to
$d_{osc}(E)$ into two parts. In the first we collect all periodic orbits whose
matrices have traces less than a certain value $n_0\gg 1$ and all others are
putted in the second part. The first one we simply add to $\tilde d(E)$ which
from now will contain all "non--interesting" terms which are explicitly known
and could be calculated without principal difficulties.
Neglecting the difference between
$2\cosh L$ and $e^L$ and further disregard the existence of multiple
traversals (since they are exponentially few in number compared to the
primitive orbits) we concentrate on

\begin{equation}
d_{osc}(E)={2\over\pi k}\sum_{n=n_0}^\infty g(n){\ln n\over n}\cos(2k\ln
n).\eqnoi\end{equation}
(All other terms are putted to $\tilde d(E)$.)

We have seen above that
the coefficient entering in Eq.~\last{} is on the average
of order one. Defining
\begin{equation}
\eqnoi
\alpha(n)=g(n)\ln n/n,
\end{equation}
we finally obtain that
\begin{equation}
d_{osc}(E)={2\over\pi k}\sum_{n=n_0}^\infty \alpha(n)\cos(2k\ln n),
\eqnoi\label{main}
\end{equation}
where $\avg{\alpha(n)}$ is one.

Eq.~\last{} as it is written diverges when the sum is performed over all
values of $n$. This is the well-known divergence of the summation over long
periodic orbits, which is inherent in all semiclassical formulae.
Mathematically, one treats such problems by using a suitable function
$h(r)$ in the Selberg trace formula. We here use a different method of
regularization based on the subtraction of the main term in the density
of periodic orbits (see e.g., \cite{{berry},{bogom},{aurich1}}). In Eq.~\last,
it corresponds to the substitution
\def\alphat{\tilde\alpha}
of $\alpha(n)$ by $\alphat(n)=\alpha(n)-1$, i.e., the subtraction of the
mean value of $\alpha(n)$\footnote{This subtraction is not necessary for
billiard problems with Dirichlet boundary conditions}.
Of course, one has to add to $\tilde d(E)$ the rest
$$\sum_{n=n_0}^{\infty}\cos(2k\ln n)=\Re\,
\zeta(2ik)-\sum_{n=1}^{n_0-1}\cos(2k\ln n),$$
where $\zeta(s)$ is the Riemann zeta function whose analytical continuation can
be done easily by using the well--known functional relation \cite{tit}.

Finally the level density of energy levels for the modular group can be written
in the following form
\begin{equation}
d(E)=\avg{d(E)}+d_{osc}(E),
\eqnoi\label{main2}
\end{equation}
where
$$d_{osc}(E)={2\over\pi k}\sum_{n=n_0}^\infty \alphat(n)\cos(2k\ln n),$$
and for simplicity we redefine $\avg{d(E)}$:
$$\avg{d(E)}=\frac{1}{6}+\tilde d(E).$$
Usually the mean density of levels includes only the Weil term (plus
corrections). Sometimes it is convenient to put to it other terms as well.
Here we shall consider all explicitly known terms and the convergent
contributions from repetitions of primitive periodic orbits as part of
$\avg{d(E)}$. They are unessential for our purposes but can be important for
numerical calculations.

Having the expression for the level density, one can formally
compute the $n$-point correlation functions:
\begin{equation}
\eqnoi
R_n(\eps_1,\ldots,\eps_n)=\avg{d(E+\eps_1)d(E+\eps_2)\cdots
d(E+\eps_n)}.
\end{equation}
The important point to note here is the energy smoothing denoted
by the brackets:
\begin{equation}
\eqnoi
\avg{f(E)}=\int f(E^\prime)\sigma (E-E^\prime)dE^\prime
\end{equation}
and $\int \sigma (x)dx=1$.
Here $\sigma(x)$ is a function which is peaked near $x=0$ and has a width
$\Delta E$.  It is usually
assumed that $\avg{d}^{-1}\ll \Delta E\ll E$, where $\avg{d}$ is the mean
level density. The standard choice is the Gaussian:
$$\sigma (x)=\frac{1}{\sqrt{2\pi}\Delta E}
\exp \left (-\frac{x^2}{2(\Delta E)^2}\right )$$
but other smoothing fuctions are also possible. They should be choosen in such
a way that if $u$ is a constant $\avg{\exp (iku)}\rightarrow 0$
sufficiently fast as $k\rightarrow \infty$.
This kind of averaging procedures is inevitable
for the statistical analysis of a system in which there are
no random parameters.

In this paper we concentrate on the two-point function
\begin{equation}
R_2(\eps_1,\eps_2)=\avg{d(E+\eps_1)d(E+\eps_2)},\eqnoi
\label{2.26}
\end{equation}
where for the modular group we use the following
expressions:
\begin{equation}
\eqnoi
d(E)=\avg{d(E)}+d_{osc}(E),
\end{equation}
where as $E \rightarrow \infty $ $\avg{d(E)} \rightarrow 1/6$
and $d_{osc}(E)$ is given by
Eq.~(\ref{main2}). One then finds
\begin{equation}
R_2(\eps_1,\eps_2)=\avg{\avg{d(E)}^2}+\overline R_2(\eps_1,\eps_2),
\eqnoi
\end{equation}
where
\begin{eqnarray}
\overline R_2(\eps_1,\eps_2)&=
&{1\over(\pi k)^2}\sum_{n_1,n_2}\alphat(n_1)\alphat(n_2)
\langle e^{2i(k_1\ln n_1+k_2\ln n_2)}\nonumber\\
& & \mbox{} + e^{2i(k_1\ln n_1-k_2\ln n_2)}+c.c.\rangle
\eqnoi
\end{eqnarray}
and $k_i$ is $\sqrt{2(E+\eps_i)}$ which goes as $k+\eps_i/k$
as $k\to\infty$.

Due to the energy average, the first term will be washed out
and the second one will give contributions only when $n_2=n_1+r$ with
$r\ll n_1, n_2$. Finally
\begin{equation}\eqnoi
\overline R_2(\eps_1,\eps_2)=
{1\over\pi^2k^2}\sum_{n=n_0}^\infty\sum_{r=-\infty}^\infty
\alphat(n)\alphat(n+r)
\left[\exp\left(2i{kr\over n}-2i\eps{\ln n\over k}\right)+\ccc\right].
\end{equation}
where $\eps=\eps_1 - \eps_2$. Note that the two-point function
depends on the difference $\eps=\eps_2-\eps_1$ as it should be.

Let us now assume that the following mean value exists
\begin{equation}
\eqnoi
\gamma(r)=\lim_{N\to\infty}{1\over N}\sum_{n=1}^N\alphat(n)\alphat(n+r).
\label{gr}
\end{equation}
Then the dominant contribution to the two-point correlation function
will be given by
\begin{equation}
\eqnoi
\overline R_2(\eps)=
{2\over\pi^2k^2}\Re\int_{n_0}^\infty dn \sum_{r=-\infty}^\infty
\gamma(r)e^{2ikr/n} \exp\left(-2i\eps{\ln n\over k}\right),
\end{equation}
where we have used a continuum approximation for $n$, since
only large values of $n$ make a significant contribution.
If we now define $f(x)$ as follows:
\begin{equation}
f(x)=\sum_{r=-\infty}^\infty \gamma(r)e^{irx},\eqnoi\end{equation}
we can finally express the two-point function as
\begin{equation}
\overline R_2(\eps)={1\over\pi^2k}\int_{\tau_0}^\infty e^{ku/2}f(2ke^{-ku/2})
\cos\eps u\,du,
\eqnoi
\end{equation}
where $\tau_0=2\ln n_0/k$.

Another frequently used quantity is its Fourier transform $K(t)$,
(the two-point form factor)
\begin{equation}
K(t)={1\over2\pi}\int_{-\infty}^\infty \overline R_2(\eps)e^{i\eps t}dt
={1\over2\pi^2k}e^{kt/2}f(2ke^{-kt/2})={1\over\pi^2w}f(w),
\label{form}
\eqnoi\end{equation}
where $w$ is $2ke^{-kt/2}$ and $K(t)=0$ when $t<\tau_0$.
Therefore all the non-trivial information
is contained in the functions $\gamma(r)$ and $f(x)$.
In the following section, we will outline a general method for evaluating
them. The simplest approximation, known as the diagonal approximation
\cite{berry}, would be to assume that the
$\alpha(n)$ are essentially uncorrelated, that is, that $\gamma(r)$
is zero for $r\neq0$. This gives for $f(x)$ a constant value, which
leads to an exponential growth of $K(t)$ as is clear from Eq.~\last
\, \cite{{georgeot},{georgeot2}}.
On the other hand, from a general consideration \cite {berry}
(see Section 5) one sees that $K(t)$
must saturate to a constant value for $t\to\infty$ if it was
originally obtained from a discrete spectrum. The exact
expression for $f(x)$ found in the next Section will give
a resolution of this discrepancy, as will be seen in great detail
in the final Sections.

As a final remark, we note that Eq.~\last{} indicates that the
modular domain in fact behaves much as an ordinary integrable system,
in spite of its chaotic classical behaviour. Indeed, as we shall see,
$K(t)$ reaches the constant value of the order of one at a
time $t^{\star}\approx \ln k/k$ which goes to zero as
$k\to\infty$. This means that in the region where the two-point correlation
function is expected to be universal, namely for distances of the order
of a mean level spacing, and hence for $t$ not excessively small, the
eigenvalues of the modular domain do not show correlation. On the other
hand, there is structure present at small times, due to the cumulative effect
of short periodic orbits.

\chapitre{Two-point Correlation Function of Multiplicities}
\setcounter{equation}{0}

In this section we shall give a way of evaluating the quantities $\gamma(r)$
and $f(x)$ defined in the previous section. The first remark
concerning the correlations $\gamma(r)$ is that they have undamped
oscillations related to their number-theoretical nature. This recalls to
some extent the correlation between prime numbers which show similar
oscillations, as shown by the Hardy--Littlewood conjecture
\cite{hardy}. We shall therefore
follow a similar line of approach to evaluate $\gamma(r)$.

The key point is to find a suitable expression for the multiplicity
of periodic orbits of the modular group. In the previous
Section we introduced the normalized multiplicities $\alpha(n)$
equal to $g(n)\ln/n$ and it was shown that $\avg{\alpha(n)}$
is equal to one.
In Fig.~2 we present the correlations of the multiplicities $\alpha(n)$,
that is, we show $\gamma(r)$. What is striking about that graph is
that it shows correlations which do not decay as $r\to\infty$. This is
in a sense similar to the situation prevailing for the correlations between
primes, as shown by the Hardy--Littlewood conjecture. We shall therefore
follow a somewhat similar path.
We define
\begin{eqnarray}
\alpha(q;r)&=&\lim_{N\to\infty}{1\over
N}\sum_{m=0}^{N-1}\alpha(mq+r)\nonumber\\
&=&\lim_{u\to0}(qu)\sum_{m=0}^\infty\alpha(mq+r)
e^{-(mq+r)u},
\eqnoi
\end{eqnarray}
where the equality between both is guaranteed by a Tauberian theorem
\cite{{tauber},{feller}}.
The intuitive meaning of this quantity is the average value of $\alpha(n)$
when $n$ only runs over numbers of the form $mq+r$ for given $q$ and $r$.
Clearly, since $\avg{\alpha(n)}$ is one
\begin{equation}
\sum_{r=0}^{q-1}\alpha(q;r)=q.\eqnoi\end{equation}
Now, if $\alpha(n)$ were a smooth function, we would expect all $\alpha(q;r)$
to be equal to each other and hence to one. This, as we shall see, is not
at all the case. Rather, the dependence of $\alpha(q;r)$ on its arguments
is exceedingly complex and highly irregular. This will in fact be one of the
principal technical difficulties of this subject.

In Appendix A, we shall show the exact way to compute $\alpha(q;r)$. To
state the result, we need a few definitions: We call $M_q$ the set
of $2\times2$ matrices with entries being integers modulo $q$ and having
determinant one modulo $q$.
These matrices form a group under multiplication modulo $q$ and
is sometimes called the modulary group \cite{rankin}.
Additionally, we define $M_{q,r}$ to be the set of
elements of $M_q$ with trace equal to $r$ modulo $q$.
We generally denote the number of elements of a set $M$ by $|M|$.
The result of Appendix A can then be stated as follows:
\begin{equation}
\alpha(q;r)={q|M_{q,r}|\over|M_q|}.\eqnoi\end{equation}
In Table 1 we present the calculated values of $\alpha(q;r)$ for $q\leq 11$.
These results are in fact in very good agreement with direct numerical
computations of $\alpha(q;r)$.
The intuitive meaning
of this apparently strange result is the following: $g(n)$ is the
number of conjugacy classes of modular matrices of trace $n$. To each
modular matrix, one can associate an element of $M_q$ in a unique way
simply by taking the entries of the matrix modulo $q$. If $n$ is equal to
$r$ modulo $q$, then all these matrices will belong to $M_{q,r}$.
If we therefore assume that the matrices of the modular
group cover the set $M_q$ in some sense uniformly,
Eq.~\last{} appears reasonable. The argument presented in Appendix A
makes these ideas more rigorous.

Now, in order to compute the correlation $\gamma(r)$ between
the $\alpha(n)$ we proceed as follows: Define as in \cite{hardy}
\begin{equation}
\Phi(z)=\sum_{n=0}^\infty \alpha(n)z^n.\eqnoi\end{equation}
Since $\avg{\alpha(n)}$ is one, the convergence radius
of this series is equal to one. The importance of this function comes
from the fact that
\begin{equation}
\eqnoi
J_r(e^{-u})=e^{ru}\int_0^{2\pi}{d\phi\over2\pi}\Phi^*\left(e^{-u+i\phi}
\right)\Phi\left(e^{-u-i\phi}\right)e^{-ir\phi}=\sum_{n=1}^\infty
\alpha(n)\alpha(n+r)e^{-2nu}
\label{hardy}
\end{equation}
and the right-hand side, again by a Tauberian theorem
\cite{{tauber},{feller}}, is connected to the quantity $\gamma(r)$
which we wish to obtain.

The essence of the Hardy--Littlewood approach \cite{hardy} is the
investigation of the function $\Phi(z)$ as $z=\exp(-u+i\eps+2\pi ip/q)$
as $u\to0$ and $\eps\to0$, where $p$ and $q$ are coprime integers.
The main step is then to write $n$ in the form $mq+r$ with $r$
lying between $0$ and $q-1$ and prove that in the expression for
$\Phi(z)$ in Eq.~\back{1} the dominant contribution as $u$ and
$\eps$ go to zero will be given by the {\em mean} value of
$\alpha(mq+r)$, that is, one may substitute it by $\alpha(q;r)$.
We present a more detailed discussion of the validity of this
assumption in Appendix B. Here it may be sufficient to say that the
basic assumption involved is the one that there are {\em no} ordinary
short-range correlations involved. Rather, all correlations
have the oscillatory long-range behaviour shown in Fig.~2.
Accepting this, one has that as $u\rightarrow 0$ and $\eps \rightarrow 0$
\begin{eqnarray}
\eqnoi
\Phi\left(\exp\left(-u+\tpipq+i\eps\right)\right)&=&\sum_{r=0}^{q-1}
\sum_{m=0}^\infty\alpha(mq+r)e^{-(u-i\eps)}e^{2\pi ipr/q}\nonumber\\
&=&\sum_{r=0}^{q-1}\alpha(q;r)e^{2\pi ipr/q}{1\over q}
\int_0^\infty dn\,e^{-(u-i\eps)n}\nonumber\\
&=&{\beta(p,q)\over u-i\eps}
\end{eqnarray}
where
\begin{equation}
\beta(p,q)=q^{-1}\sum_{r=0}^{q-1}\alpha(q;r)\exp\left(\tpipq
r\right),\eqnoi\end{equation}
Therefore the function $\Phi(z)$ has a pole singularity at
all points on the unit circle which have a rational multiple
of $2\pi$ as phase. Its explicit form is given in Eq.~\back{1}.
The reason for the appearance of such poles is the irregular
behaviour of $\alpha(n)$ and, connected with it, the fact that
the $\alpha(q;r)$ depend in a highly non-trivial way on $q$ and $r$
instead of being independent of them, as would be the case for
more ``reasonable'' sequences.

The next step is to substitute Eq.~\back{1} into Eq.~\back{2}
We therefore
divide the unit circle in intervals $I_{p,q}$ centered around
$\exp(2\pi ip/q)$, where $p$ and $q$ run over all relatively prime numbers
with $p<q$ and $q$ less than some prescribed upper bound $Q$ which
later goes to infinity. If one now divides the integral in this way,
neglects all terms in each interval except the pole terms of Eq.~\back{1}
and finally extends the integration over $\eps$ to the whole line, one
obtains:
\begin{eqnarray}
\eqnoi
J_r(e^{-u})&=&e^{ru}\sum_{(p,q)=1}\int_{-\infty}^\infty{d\eps\over2\pi}
{|\beta(p,q)|^2\over u^2+\eps^2}e^{ir(2\pi p/q+\eps)}\nonumber\\
&=&{1\over2u}\sum_{(p,q)=1}|\beta(p,q)|^2\exp\left(\tpipq r\right)
\end{eqnarray}
as $u\to0$. Here and in the following, $(p,q)$ will denote the
greatest common divisor of $p$ and $q$. Finally, from
the definition of $J_r(e^{-u})$ in Eq.~\back{3} one obtains for
$\gamma(r)$
\begin{equation}
\eqnoi
\gamma(r)=\sum_{(p,q)=1}|\beta(p,q)|^2\exp\left(\tpipq r\right).
\end{equation}
The sum is performed over all $q$, all $0<p<q$ and coprime to $q$,
and the term $p=0$ and $q=1$ is omitted as we defined $\gamma(r)$ through
$\alphat(n)$ (see \ref{gr}) whose mean value is zero.

This is the two-point correlation function of the multiplicities
of the periodic orbits for the modular group . All other quantities
of interest can be obtained from it. In particular the function
$f(x)$ introduced in the previous Section is given by
\begin{equation}
\eqnoi
f(x)=2\pi\sum_{(p,q)=1}|\beta(p,q)|^2\delta(x-2\pi p/q),
\end{equation}
where the summation is over all $p$ and $q$ coprime, without
the restriction $p<q$.

The subtraction needed to make the trace formula converge (see Section 2)
is equivalent to setting $f(0)$ equal to zero which is
equivalent to removing from the sum in eq.~\last{} those
terms for which $p=mq$, where $m$ is an integer. From now on we
assume implicitly that these terms are removed from the sum and the
``renormalized'' $f(x)$ satisfies $f(0)=0$.
As we saw above, the knowledge of $f(x)$
determines immediately the function $K(t)$
and consequently its Fourier transform, which is the two-point function.

This is in essence the main result of our paper. Combined with a
fairly technical evaluation of $\beta(p,q)$ which is carried out
in Appendices C and D, this  gives a closed form for the Fourier
transform of the two-point function which is presumably exact. Note
that in these evaluations there are no (or merely
trivial) approximations. The substantial problems arise from the
approximations involved in the use of the Hardy--Littlewood
method as well as in the various simplifications of the
Selberg trace formula.

\chapitre{Results}
\setcounter{equation}{0}

The fundamental equation worked out at the end of the preceding section
gives an exact form for the two-point function as well as its
Fourier transform, but they are still somewhat unwieldy. To simplify
them and cast them in a useful form will be the purpose of this Section.

To this end we need to point out a basic number-theoretical property
of the functions we have been discussing: they are all so-called
{\it multiplicative\/} functions. One says that the function
$g(n)$ is multiplicative when it has the following property
\begin{equation}
g(mn)=g(m)g(n)\qquad\hbox{whenever $(m,n)=1$}.\eqnoi\end{equation}
Multiplicative functions are therefore uniquely determined by their
values on numbers of the form $p^\alpha$ where $p$ is a prime number.
It turns out that $\alpha(q;r)$ is multiplicative in the argument $q$
and the functions $\beta(p,q)$ as well. The first follows from the
fact that $M_{q,r}$ and $M_q$ can be expressed as the number of solutions
of a given set of congruences modulo $q$. These are therefore multiplicative
in $q$ as a consequence of the following well-known fact known as the
Chinese Remainder Theorem(see e.g. \cite{basic}):
If $q_1$ and $q_2$ are relatively prime,
then to every solution of a congruence or set of congruences modulo
$q_1q_2$ there corresponds uniquely a pair of solutions to the
same congruences mod $q_1$ and $q_2$ respectively and vice versa.
The fact that more complicated expressions such as $\beta(p,q)$ maintain
this multiplicative property is a tedious but straightforward exercise.

To simplify the expression for $f(x)$, we will use the following
identity valid for an arbitrary multiplicative function $g(n)$:
\begin{equation}
\sum_{n=1}^\infty g(n)=\prod_{p{\rm\ prime}}(1+\sum_{k=1}^\infty
g(p^k)),\eqnoi\end{equation}
which is known as Euler's identity. This leads to
\begin{eqnarray} \eqnoi
\gamma(r)&=&\sum_{n=1}^\infty A_r(n)-1\nonumber\\
&=&\prod_p (1+\sum_{k=0}^\infty A_r(p^k))-1,
\end{eqnarray}
where $A_r(q)$ is given by
\begin{equation}
A_r(q)=\sum_{p:(p,q)=1}\left|\beta(p,q)\right|^2\exp\left(\tpipq r
\right).\eqnoi\end{equation}
To give a closed expression for $A_r(q)$ we still need one standard
definition from number theory: we define the Legendre symbol
of $a$ and $q$, where $q$ is an odd prime as follows
\begin{eqnarray} \eqnoi
\jac{a}{q}&=&1\qquad\hbox{if there is an $x\not\equiv0\pmod q$
such that $a\equiv x^2\pmod q$,}\nonumber\\
&=&0\qquad\hbox{if $a\equiv0\pmod q$}\\
&=&-1\qquad\hbox{otherwise.}\nonumber
\end{eqnarray}
The meaning of this
number is perhaps best understood by saying that the number of
{\it distinct\/} solutions of the equation $x^2\equiv a\pmod q$
is $1+(a/q)$.
The properties of the Legendre symbol are stated in any standard
reference on number theory (see e.g. \cite{{basic},{vino}}).
A fairly tedious evaluation of $A_r(q)$ (performed in Appendix D)
gives the following for $q$ a prime power (which is all that is necessary,
since $A_r(q)$ is also multiplicative in $q$): First, let
$q$ be equal to $p^n$, where $p$ is an odd prime. Then we have for $n=1$:
\begin{equation}
A_r(p)={1\over(p^2-1)^2}\left[p\sum_{x=0}^{p-1}
\jac{(x^2-4)((x+r)^2-4}{p}-1\right].
\label{oddpr}
\eqnoi\end{equation}
For $n\geq2$, we have, letting $t$ be an arbitrary non-zero
number modulo $p$
\begin{equation}
\eqnoi
A_r(p^n)={1\over p^{2n}(1-p^{-2})}\left\{
\begin{array}{ll}
2(1-1/p)&r\equiv0\pmod{p^n}\\
-2/p&r\equiv tp^{n-1}\pmod{p^n}\\
\eps(n,p)(1-1/p)&r\equiv\pm4\pmod{p^n}\\
-\eps(n,p)/p&r\equiv\pm4+tp^{n-1}\pmod{p^n}
\end{array}
\right.
\label{oddprpower}
\end{equation}
where $\eps(n,p)$ takes the value $-1$ if $n$ is odd and $p$ is
of the form $4k+3$ and is equal to $1$ in all other cases.
For $p=2$, we must list down individual cases for low powers
and eventually state a general rule:
\begin{eqnarray}
\eqnoi\label{2prime}
A_r(2)&=&\left\{
\begin{array}{ll}
1/9&r\equiv0\pmod2\\
-1/9&r\equiv1\pmod2
\end{array}\right.\\
\eqnoi
A_r(4)&=&\left\{
\begin{array}{ll}
1/18&r\equiv0\pmod4\\
-1/18&r\equiv2\pmod4
\end{array}\right.\\
\eqnoi
A_r(8)&=&0\\
\eqnoi
A_r(16)&=&1/(9\cdot16)\left\{
\begin{array}{ll}
1&r\equiv0\pmod{16}\\
-1&r\equiv8\pmod{16}
\end{array}\right.\\
\eqnoi
A_r(32)&=&0
\end{eqnarray}
and finally, for the general case $n\geq 6$
\begin{equation}
A_r(2^n)=1/(9\cdot2^{2n-4})\left\{
\begin{array}{ll}
2&r\equiv0\pmod{2^n}\\
-2&r\equiv2^{n-1}\pmod{2^n}\\
1&r\equiv\pm(4+2^{n-2})\pmod{2^n}\\
-1&r\equiv\pm(4+2^{n-2}+2^{n-1})\pmod{2^n}\\
\end{array}\right.
\label{2power}
\eqnoi
\end{equation}
All the terms not explicitly shown above are of course equal to zero.

\chapitre{Two-point Form Factor of Energy Levels}
\setcounter{equation}{0}

We have shown in Section 2 that the Fourier transform of the
two-point correlation function of the energy levels for the
modular group can be written as follows:
\begin{equation}
\eqnoi
K(t)=\frac{1}{\pi^2w}f(w),
\end{equation}
where $w=2ke^{-k|t|/2}$ and
\begin{equation}
\eqnoi
\label{5.2}
f(x)=2\pi\sum_{(p,q)=1}|\beta(p,q)|^2\delta(x-2\pi p/q),
\end{equation}
and $\beta(p,q)$ can be expressed in terms of Kloosterman
sums as in Eq.~(\ref{kloos}).

The two-point correlation function of any system with a non--degenerate
discrete spectrum should have the following asymptotic behaviour \cite{berry}:
\begin{equation}
\eqnoi
R_2(\eps_1,\eps_2)\to\avg{d(E)}\delta(\eps_2-\eps_1)\qquad\hbox{as
$\eps_2-\eps_1\to0$}
\end{equation}
This is a simple consequence of the fact that, in the absence
of systematic level degeneracies in the sum
\begin{equation}
\eqnoi
R_2(\eps_1,\eps_2)=\sum_{n_1,n_2}\delta(E-E_{n_1}+\eps_1)
\delta(E-E_{n_2}+\eps_2)
\end{equation}
only the terms with $E_{n_1}=E_{n_2}$ are important in the limit
$\eps_2\to\eps_1$. From this the asymptotic relation of Eq.~\back{1}
follows. Consequently the two-particle form factor $K(t)$ should
have the following large $t$-asymptotics:
\begin{equation}
\eqnoi
K(t)\to{1\over2\pi}\avg{d(E)}\qquad\hbox{as $t\to\infty$}
\end{equation}
(For the modular domain $\avg{d(E)} \rightarrow 1/6$ as $E\rightarrow \infty$.)
For the Poisson distribution,
the form factor $K(t)$ takes its asymptotic value throughout the range
of $t$ of order one (there is a non-universal region near $t=0$).
In the case of the standard matrix ensembles (GOE, GUE and GSE)
these form factors reach their asymptotic value when $t\gg1$.

Although the above limiting behaviour follows from quite
general considerations, it is generally not possible to
derive it from the semiclassical formulae. Indeed, this fact depends
on interference between non-degenerate  long periodic orbits
and is therefore quite difficult to ascertain. In the case
of the modular domain, our knowledge is by now sufficiently
detailed that we can attempt an explicit verification of
this behaviour. Of course, since the Selberg trace formula
is exact and the spectrum of the modular domain is discrete,
there is every reason to expect that it will work.
Nevertheless, such a calculation provides an extremely powerful
consistency check on the various approximations made in order
to arrive at the final result. This is what we shall do in this
Section.

The ``exact'' form factor as given in Eq.~(\ref{form}) is really a
sum of delta functions at the points
\begin{equation}t_{p,q}={2\over k}\ln{kq\over\pi p}.\eqnoi
\end{equation}
Eq.~(\ref{form}) can then be rewritten as
\begin{equation}
K(t)={1\over\pi^3k}\sum_{(p,q)=1}\left|{q\over p}\beta(p,q)\right|^2
\delta(t-t_{p,q}).\eqnoi
\end{equation}
The limit $t\to\infty$ then corresponds to the behaviour of $f(x)$
as $x\to0$.

In Fig.~3a we present a plot of $f(x)$ computed from the periodic orbits of
the modular domain using a direct Fourier transformation of the
computed $\gamma(r)$ with an appropriate smoothing procedure
\begin{equation}
\tilde f(x)=\frac{1}{N}\sum_{k=0}^N\gamma(k)e^{-\lambda k^2}\cos kx,
\eqnoi
\end{equation}
where the smoothing parameter $\lambda$ has been taken equal to $3/N^2$
and $N=1000$. The big peaks corresponding to small values of $p$
and $q$ are well pronounced, but its behaviour between peaks
is not clear. It can be visualize if instead of $f(x)$ one computes its
integral
\begin{equation}
G(x)=\int_{0}^{x} f(y)dy.
\eqnoi
\label{Gx}
\end{equation}
The plot of $G(x)$ is given in Fig.~3b. These pictures suggest that it is
convenient to separate the function $f(x)$ into two parts
\begin{equation}
f(x)=f_N^{osc}(x)+f_N(x),
\eqnoi
\end{equation}
where the first function when $0<x<2\pi$ is given by Eq.~(\ref{5.2})
but with finite number of terms for which $q\leq N$
\begin{equation}
f_N^{osc}(x)=2\pi\!\!\sum_{
\begin{array}{c}
{\scriptstyle(p,q)=1}\\
\scriptstyle0<p<q\leq N
\end{array}
}\!\!|\beta(p,q)|^2\delta(x-2\pi p/q),
\eqnoi
\label{fosc}
\end{equation}
and the second function $f_N(x)$ includes contributions
from all terms with  $q>N$
\begin{equation}
f_N(x)=2\pi\!\!\sum_{
\begin{array}{c}
{\scriptstyle(p,q)=1}\\
\scriptstyle q>N
\end{array}
}\!\!|\beta(p,q)|^2\delta(x-2\pi p/q),
\eqnoi
\end{equation}
The calculation of the first function is straightforward and only the second
one needs a special attention.

Of course, one can simply calculate $f^{osc}_{N}(x)$ for sufficiently
large $N$  but we shall show that adding to it an approximate expression for
$f_N(x)$ leads to
much better approximation even for small $N$.

It is shown in Appendix C that $|\beta(p,q)|\rightarrow 0$ as $q\rightarrow
\infty$. Therefore all peaks in $f_N(x)$ are small and it is resonable to
compute its mean value. We shall proceed
in the following way (see e.g. \cite{basic}). Let us define the function
\begin{equation}g_N(x)=\int_0^x f_N(y)dy.\eqnoi
\end{equation}
If we now find a smooth function $G_N(x)$ such that
\begin{equation}|G_N(x)-g_N(x)|\ll G_N(x),\eqnoi
\end{equation}
then we shall say that $g_N(x)$ and $G_N(x)$ are of the same order. We may then
define
\begin{equation}\avg{f_N(x)}={dG_N(x)\over dx}.\eqnoi
\end{equation}
Note that this method is equivalent to smoothing the form factor over a small
energy interval as in Eq.~(\ref{2.26}).

Formally:
\begin{equation}
g_N(x)=2\pi\!\!\sum_{
\begin{array}{c}
{\scriptstyle(p,q)=1}\\
{\scriptstyle p/q <x/(2\pi)}\\
\scriptstyle q>N
\end{array}
}\!\!|\beta(p,q)|^2,
\eqnoi
\label{5.12}
\end{equation}
where the summation is over all coprime integers such that
$$0<p/q<{x\over 2\pi},\;\;q>N$$
When $q$ is fixed, $\beta(p,q)$ is a number-theoretical function
of $p$. It is in fact quite erratic. Therefore we conjecture that it
can to some degree of approximation be replaced by its mean value
over all $p$, which we call $\beta(q)$:
\begin{equation}\beta(q)=
\avg{|\beta(p,q)|^2}={\sum_{(p,q)=1}|\beta(p,q)|^2\over
\sum_{(p,q)=1}1},\eqnoi
\end{equation}
where the summation is performed over all $p$ coprime to and less
than $q$. But it is easy to see that
\begin{equation}\sum_{(p,q)=1}|\beta(p,q)|^2=A_0(q)=\prod_i
A_0(\omega_i^{n_i}),\eqnoi
\end{equation}
where the $\omega_i$ are the prime factors of $q$ and $n_i$ the power
with which they occur. These values have already been computed in Appendix D.
It is worth pointing out that in the case of an odd prime one can
reduce the intractable expression of Eq.~(\ref{oddpr}) to the following:
\begin{equation}
A_0(p)={p^2-2p-1\over p^4(1-p^{-2})^2}. \eqnoi
\end{equation}

Further, since one has, by elementary number theory
\cite{{basic},{vino}}, the following
facts about the Euler function
\begin{equation}\varphi(q)=\sum_{(p,q)=1}1=\prod_{\omega|q}
\left(1-{1\over\omega}\right),\eqnoi
\end{equation}
it follows that $\beta(q)$ is known explicitly (see Appendix F).
Let
\begin{equation}
G_N(x)=2\pi\!\!\sum_{
\begin{array}{c}
{\scriptstyle(p,q)=1}\\
{\scriptstyle 0<p/q<x/(2\pi)}\\
\scriptstyle q>N
\end{array}
}\beta(q).\eqnoi
\label{5.18}
\end{equation}
The calculation of this function is performed in Appendix F. Here for clarity
we consider a simple prototype of such function.

Let
\begin{equation}
j_N(x)=\sum_{
\begin{array}{c}
{\scriptstyle 0<p/q<x}\\
\scriptstyle q>N
\end{array}
}\!\!\frac{1}{q^3}.
\eqnoi
\end{equation}
It differs from the exact function (\ref{5.18}) in two things. First, the sum
includes all $p$ and not only coprime to $q$. Second, the function $\beta(q)$
is substituted by its asymptotic behaviour. Now
\begin{equation}
j_N(x)=\sum_{q=N+1}^{\infty}\frac{[xq]}{q^3},
\eqnoi
\end{equation}
where $[y]$ is an integer part of $y$.

When $Nx\ll 1$ and $N\rightarrow \infty$ one can replace the sum by an
integral and one obtains
\begin{eqnarray}
j_N(x)&\approx& x^2\int_0^\infty\frac{dq\,[q]}{q^3}
\nonumber\\
&=&{x^2\over2}\sum_{n=1}^\infty{1\over n^2}
\nonumber\\
&=&\frac{\pi^2}{12}x^2.
\eqnoi
\end{eqnarray}
When $Nx\gg1$, one has $[xq]\approx xq$ and
\begin{equation}
j_N(x)\approx x\sum_{q=N+1}^{\infty} \frac{1}{q^2}=\frac{x}{N}.
\eqnoi
\end{equation}
Therefore at small $x\ll 1/N$ $j_N(x)$ behaves as $\pi^2 x^2/12$ but at
$x\gg1/N$ ($x<2\pi$) it grows as $x/N$. The exact function (\ref{5.18}) has the
similar behaviour but the computations are more complicated. The details are
presented in Appendix F.

When $x\ll 1/N$
\begin{equation}
G_N(x)=\frac{\pi}{24} x^2,
\eqnoi
\end{equation}
Therefore one has
\begin{equation}
\eqnoi
\avg{f_N(x)}={\pi x\over12}\qquad(x\to0).
\label{5.25}
\end{equation}
When $x\gg 1/N$ and $(2\pi - x)\gg 1/N$ the asymptotics of $G_N(x)$ changes. In
this region
\begin{equation}
G_N(x)\rightarrow \frac{C}{N}x,
\label{5.26}
\eqnoi
\end{equation}
where
$$C=\prod_{p}(1-\frac{1}{p(p+1)})$$
and the product is taken over all primes including $p=2$. Numerically
$C\approx .704$.

 Combining these values one concludes that the function $\avg{f_N(x)}$ as
$0<x<2\pi$ approximately has the shape as in Fig.~4 and continues periodically
beyond this interval.

In our approximation the function $f(x)$ can be written in the simple form:
$$f(x)=f^{osc}_{N}(x)+\avg{f_{N}(x)}.$$
Together with the explicit
expression for $f_N^{osc}(x)$ in Eq.~(\ref{fosc}) it gives a quite accurate
description of the two--point form factor for the modular domain.

In Fig.~5 we presented the difference between the `exact' function $G(x)$
computed by taking into account all terms with $q\leq 1000$ and the sum of
these two terms for different value of $N$. Note the difference in scales with
respect to Fig.~3. Quite good agreement is observed even for $N=20$.

The knowledge of the function $\avg{f_N(x)}$ is particular important in the
region of small $x$ because it gives the dominant contribution to the
asymptotics of the two--point correlation formfactor. From Eqs.~(\ref{5.25})
and (\ref{5.26}) it follows that as $x\rightarrow 0$
\begin{equation}
f(x)\rightarrow \frac{\pi}{12}x\;\; \mbox{and} \;\;
G(x)\rightarrow \frac{\pi}{24}x^2.
\label{as}
\eqnoi
\end{equation}
In Fig.~6 we plot the function $G(x)$ computed as the sum of all terms
up to $N=500$ and $N=1000$ in the double logarithmic scale which is the most
sensitive to small $x$ behaviour. The solid line is the asymptotics (\ref{as}).

The most important consequence is that in the limit $k\rightarrow \infty$
and $t$ fixed the two-point formfactor tends to the constant value
\begin{equation}
K(t)=\frac{1}{12\pi},
\eqnoi
\end{equation}
as it should be for the Poisson distribution.

\chapitre{Billiard problems}
\setcounter{equation}{0}

We have mentioned that eigenfunctions of the Laplace--Beltrami operator
for the modular group can be classified by the parity with respect to the
inversion $x\rightarrow -x$. The odd (even) functions are eigenfunctions of the
billiard problem with the Dirichlet (Neumann) conditions on the boundary
of half of the modular domain (see Fig.1). In this section we compute the
two-point correlation functions for these problems separately.

In the modular billiard problem group matrices are $2\times2$
matrices with integer
entries but with the determinant equals both $1$ and $-1$ (see e.g.
\cite{balvor}). Matrices with determinant $-1$ describe geometrically an
inversion with respect to a circle and they correspond to the following
transformation:
\begin{equation}
\pr{z}=\frac{az^*+b}{cz^*+d},
\eqnoi
\end{equation}
where $z^*$ is the complex conjugate of $z$.

The periodic orbits of the billiard problems can be identified with classes of
conjugated matrices (both with determinants $\pm 1$)
but their length is given by
Eq.~(\ref{2.5}) only if the matrix determinant equals $1$. If it equals $-1$
the length of corresponding periodic orbits should be computed by
\begin{equation}
2\sinh(l/2)=|Tr M|.
\eqnoi
\end{equation}

Physically matrices with determinant $+1$ ($-1$) correspond to periodic orbits
with even (odd) number of reflections from the billiard boundary.

For billiard problems there exist an exact Selberg--type trace formula which
expresses
the density of states through periodic orbits \cite{venkov,balvor}. It
looks like the usual trace formula (\ref{2.12}) but periodic orbits
with odd number of bounces have an additional minus sign
for Dirichlet boundary conditions.

Performing the same steps as in Section 2 one obtains:
\begin{equation}
d_{bil}(E)=\avg{d(E)}+\tilde{d}(E) + d_{osc}(E),
\eqnoi
\end{equation}
where $\avg{d(E)}=A/2\pi$ is a smooth part of the level density. $A$ here is
the area of the modular billiard equals a half of that of the modular domain:
$A=\pi/6$,
\begin{equation}
d_{osc}(E)={1\over\pi k}\sum_{n=n_0}^\infty (a^{(+)}(n)+\eps a^{(-)}(n))
\cos(2k\ln n).
\eqnoi
\end{equation}
$a^{(\pm)}(n)$ here are the normalized multiplicities of periodic orbits
corresponding to classes of conjugated matrices with determinants +1 and
$-1$
correspondingly
\begin{equation}
a^{(\pm)}(n)=2g^{(\pm)}(n)\frac{\ln n}{n},
\eqnoi
\end{equation}
where $g^{(\pm)}(n)$ is the number of periodic orbits with determinant $\pm 1$
and trace equals $n$. The factor 2 is introduced for the convenience.
$\eps =-1$ for the Dirichlet boundary conditions and $\eps =1$ for the
Neumann ones. The function $\tilde{d}(E)$ contains all other terms.

The arithmetic nature of the modular billiard group leads to the conclusion
that
\begin{equation}
\avg{a^{(\pm)}(n)}=\lim_{N\rightarrow \infty} \frac{1}{N} \sum_{n=1}^{N}
a^{(\pm)}(n) = 1.
\eqnoi
\end{equation}
Similarly to Appendix A one concludes that quantities
\begin{equation}
a^{(\pm)}(q,r)=\lim_{N\rightarrow \infty} \frac{1}{N} \sum_{m=0}^{N-1}
a^{(\pm)}(mq+r)
\eqnoi
\end{equation}
are equal to
\begin{equation}
a^{(\pm)}(q,r)=\frac{q M^{(\pm)}_{q,r}}{|M_{q}|},
\eqnoi
\end{equation}
where $M^{(\pm)}_{q,r}$ is the number of matrices
with integer entries modulo $q$
whose determinant equals $\pm 1$ and whose trace equals $r$, $|M_{q}|$ is the
total number matrices modulo $q$ with determinant 1. (The total number of
matrices with determinant $-1$ will be the same.)

After merely rephrasing arguments of previous Sections we find that the
two-point correlation form factor of the modular billiard can be written in the
following form:
\begin{equation}
K(t)=\frac{1}{4\pi^2 w}(f^{(++)}(w)+2f^{(+-)}(w)+f^{(--)}(w)),
\eqnoi
\end{equation}
where $w=2k\exp (-kt/2)$ and
\begin{equation}
f^{(\eps_1,\eps_2)}(x)=2\pi\sum_{(p,q)=1}\beta^{(\eps_1)}(p,q)
\beta^{(\eps_2)}(p,q)
\delta(x-2\pi\frac{p}{q}),
\eqnoi
\end{equation}
where
$$ \beta^{(\eps)}(p,q)=\frac{1}{q}\sum_{r=0}^{q-1} a^{(\eps)}(q,r)
\exp\left(\tpipq r\right)
$$
The explicit formulas for $\beta^{(-)}(p,q)$ are given in Appendix C.

To compute the average behaviour of $K(t)$ we have to know the average values
of $f^{(--)}(x)$ and $f^{(+-)}(x)$ as $x\rightarrow 0$.
As in the previous Section
one should first find the mean values of the product of two
$\beta^{(\eps)}$ over
all values of $p$:
\begin{equation}
\beta^{(\eps_1,\eps_2)}(q)=\frac{1}{\varphi (q)}\sum_{p:(p,q)=1}
\beta^{(\eps_1)}(p,q) \beta^{(\eps_2)}(p,q).
\eqnoi
\end{equation}
The later sum is connected to functions $A^{(\eps_1,\eps_2)}_{0}(q)$ defined in
Appendix D. If $q= \omega_1^{n_1}\omega_2^{n_2}\ldots \omega_k^{n_k}$ is the
canonical representation of $q$ as the product of different primes $\omega$
then
\begin{equation}
\beta^{(\eps_1,\eps_2)}(q)=\frac{1}{\varphi(q)}\prod_{\omega_i|q}
A^{(\eps_1,\eps_2)}_{0}(\omega_i^{n_i}).
\eqnoi
\label{beta}
\end{equation}
Using this value it is shown in Appendix F that as $x\rightarrow 0$
\begin{equation}
\label{ff}
\avg{f^{(--)}(x)}=\avg{f^{(++)}(x)} =\frac{\pi}{12}x,\;\;
\avg{f^{(+-)}(x)} =O(x^{3/2}).
\eqnoi
\end{equation}
Therefore as $t\gg \ln k/k$
\begin{equation}
K(t)\rightarrow \frac{1}{24\pi},
\eqnoi
\end{equation}
which coincides with the Poisson value for this
quantity ($K(t)=A/(2\pi)^2$ and
$A=\pi/6$).

Note that the last of relations (\ref{ff}) means that in the universal limit
eigenvalues of different symmetry classes (odd -- even with respect to the
inversion) are uncorrelated. This property is usually
taken for granted on
``general considerations''. But the problem does not seem to be trivial because
the same periodic orbits enter the trace formulas for both odd and even states.
Only their phases are different. A priori
it is unclear that the cross term will
vanish. To our knowledge the modular billiard is the only dynamical system
where the absence of correlation between states of different symmetry
can be checked analytically. (On this subject see also \cite{bogl}.)

At small values of $t\sim \ln k/k$ the billiard form factor has peaks
similar to the ones already found
in the case of the full modular group. Note that the
largest peak with $p/q=1/2$ is absent in the billiard with the Dirichlet
boundary conditions because $-1\equiv 1$ mod 2.

\chapitre{Concluding Remarks}
\setcounter{equation}{0}

In this paper we have computed the two--point correlation function for the
energy levels of the modular group and the modular billiard. From the point of
view of classical motion these systems are ergodic with strong chaotic
properties. But their arithmetical nature leads to a very large degeneracy of
lengths of periodic orbits and, as a consequence, to the fact that their
two--point correlation function tends to the Poisson value typical for the
integrable systems and not for chaotic ones. At large scale the two--point
correlation function has prominent number-theoretical oscillations.

To clarify the (tedious) derivation we very briefly repeat the essential steps.
\begin{itemize}
\item The Selberg trace formula allows
to write the density of states as
a sum over classical periodic orbits (the function $\avg{d(E)}$ is
explicitly known):
\begin{equation}
d(E)=\avg{d(E)} + \frac{2}{\pi k} \sum_{n=n_0}^{\infty} \tilde{\alpha}(n)
\cos (2k\ln n),
\eqnoi
\end{equation}
where $\tilde{\alpha}(n)=\alpha(n)-1$,
$$\alpha(n)=g(n)\frac{\ln n}{n}$$
and  $g(n)$ is the number of periodic orbits with trace equals $n$.
\item The two--point correlation form factor can be expressed as follows:
\begin{equation}
K(t)=\frac{1}{\pi^2w}f(w)
\eqnoi
\end{equation}
where $w=2k\exp(-kt/2)$,
$$f(x)=\sum_{r=-\infty}^{+\infty}\gamma (r) e^{\imath r x},$$
and
$$\gamma(r)=\lim_{N\rightarrow \infty}\frac{1}{N}\sum_{n=1}^{N}
\tilde{\alpha}(n)\tilde{\alpha}(n+r)$$
is the two--point correlation function for
the multiplicities of periodic orbits.
\item Using a probabilistic approach we show that
\begin{equation}
\alpha(q;r)=\lim_{N\rightarrow \infty}\frac{1}{N}\sum_{m=0}^{N-1} \alpha
(qm+r)=\frac{q\mqr}{|M_{q}|},
\eqnoi
\end{equation}
where $\mq$ is the total number of matrices with entries
which are taken as integers modulo $q$ and
$\mqr$ is the number of such matrices having trace equal to $r$
modulo $q$.
\item A generalization of the Hardy--Littlewood method permits to find an
explicit expression for $\gamma (r)$:
\begin{equation}
\gamma(r)=\sum_{(p,q)=1} |\beta(p,q)|^2 \exp (2\pi i\frac{p}{q}r),
\eqnoi
\end{equation}
where the summation is taken over all $q$ and all $p<q$ coprime to $q$ and
\begin{equation}
\beta(p,q)=\frac{1}{q}\sum_{r=0}^{q-1}\alpha(q;r) \exp (2\pi i
\frac{p}{q}r).
\eqnoi
\end{equation}
\item Introducing the Kloosterman sums
$$S(n,m,c)=\sum_{d=0}^{c-1}\exp\left({2\pi i\over c}(nd+md^{-1})\right),$$
$\beta(p,q)$ can be written as follows:
\begin{equation}
\beta(p,q)=\frac{1}{q^2\prod_{\omega |q} (1-\omega^{-2})}S(p,p;q),
\eqnoi
\end{equation}
where $\omega$ are the prime divisors of $q$.
\item These formulae give the explicit expression for the two--point
correlation form factor
\begin{equation}
K(t)={1\over\pi^3k}\sum_{(p,q)=1}\left|{q\over p}\beta(p,q)\right|^2
\delta(t-t_{p,q}).\eqnoi
\end{equation}
where
$$t_{p,q}={2\over k}\ln{kq\over\pi p}.$$
In the limit $k\rightarrow \infty$ and $t$ fixed, the dominant contribution
comes from terms with $p/q\ll 1$.
Smoothing over such values we show that in  this limit
$K(t)$ has the constant Poisson value:
\begin{equation}
K(t)=\frac{A}{(2\pi)^2}.
\eqnoi
\end{equation}
Here $A=\pi/3$ is the area of the fundamental region of the modular group.
For small $t$ (of order of $\ln k/k$) $K(t)$ has number--theoretical
oscillations due to cumulative contributions of degenerate periodic
orbits. For very small values of $t$ (of order of $1/k$) the two point
form factor has $\delta$ function peaks  connected with short periodic orbits.
\end{itemize}

Analogous formulas can be obtained for the modular billiards. As a byproduct
we proved that the energy levels of different symmetry are uncorrelated. It
seems that practically all such formulae can be generalized
to other arithmetical
groups. Though the modular group is by no means a generic system, it is
the first
ergodic dynamical system for which it is possible to compute explicitly the
distribution of the energy levels.

{\bf Acknowledgements}

We would like to acknowledge O. Bohigas for useful discussions.
One of the authors (F.L.) gratefully acknowledges the warm hospitality
during his stay at the IPN Orsay as well as the financial support
of DGAPA project IN 100491.

\pagebreak

\appendice{A}
\renewcommand{\theequation}{\appnum.\the\eqnum}
\setcounter{equation}{0}

Here we want to show how the expression for $\alpha(q;r)$ is derived.
One notes first that the modular group is  generated by two
elements $s$ and $t$ defined as follows:
\begin{equation}
s=\left(\matrix{0&1\cr-1&0\cr}\right)\qquad
t=\left(\matrix{1&1\cr-1&0\cr}\right).\eqnoa
\end{equation}
Since $s^2=t^3={\bf1}$, any element of the modular group can be represented
as a sequence of alternating $s$ and $t^\sigma$ where $\sigma=\pm1$.
Further, such representations are in fact unique. From this follows
that to each conjugacy class there corresponds uniquely a word
beginning with $s$ and ending with a $t^\sigma$, up to cyclic
permutations. That is to say, cyclically equivalent words generate
identical conjugacy classes, but no other words do. To this end, it
is essential, however, to enforce the condition that the word begin
with an $s$ and end with a $t^\sigma$. In the following we will
therefore only consider words generated by the matrices
\begin{equation}
m_1=st\qquad m_2=st^{-1}.\eqnoa\end{equation}
Let us now define $\alpha(q;r)$ more precisely: Up to a normalization
factor of $q$, it is the probability
that a conjugacy class belonging to a given
value of $n$ should have a trace equal to $r$ modulo $q$, after averaging
over $n$. Here instead of averaging over $n$, we shall average over
all conjugacy classes which require $k$ symbols to generate and
eventually consider the average over many and sufficiently large
values of $k$. It seems highly reasonable to claim
that these two averages should be equivalent. This is in
fact the major assumption in this Appendix.

It is obvious that there are exactly $2^k$ different words
generated by $k$ symbols. Each of those words denotes a different
conjugacy class, up to cyclic invariance. For the overwhelming majority
of words, this simply introduces a fixed factor of $k$ which does not matter
in this discussion. Those words for which the degeneracy factor is different
from $k$ are exponentially rare and can be neglected.
We can thus view the words (or conjugacy classes)
as arising from the following
Markov process: Starting from the identity, at each step multiply
the matrix to the right
randomly either by $m_1$ or by $m_2$ with probability
$1/2$. In this way
all conjugacy classes are generated with equal probability.
If one now projects this Markov process down onto the set $M_q$
by simply considering the entries of each matrix as integers modulo $q$,
then the Markov process described above projects to a Markov chain on
a finite space, namely $M_q$. Under such circumstances, very general theorems
(\cite{feller}) guarantee the existence of an approach to equilibrium. For this
purpose, one needs to show that the process is ergodic, that is,
that every element of $M_q$ can indeed be reached by an appropriate
combination of $m_1$ and $m_2$. This is seen as follows: Consider
an arbitrary element of $M_q$ as an element of the modular group.
As such, it can be represented as a product of $s$ and $t^\sigma$, though
not necessarily of $m_1$ and $m_2$. To obtain the latter, note that
there is a number $k$ such that
\begin{equation}
(st)^k={\bf1}\pmod q.\eqnoa\end{equation}
In fact $k$ can be chosen equal to $q-1$. Inserting this representation
of the identity either before or after the word, one can always bring
it into the required form.

As a final point, we need to show that the equilibrium attained on $M_q$ is
indeed the uniform distribution. This is seen by noting that $m_1$
and $m_2$ are always different and invertible, so that the Markov process
always connects an arbitrary matrix $a$ with two different matrices
$a_1$ and $a_2$ with probability $1/2$. Thus the probability of
finding matrix $a$ after $k$ steps satisfies the equation
\begin{equation}
%% FOLLOWING LINE CANNOT BE BROKEN BEFORE 80 CHAR
P_k(a)={1\over2}\left(P_{k-1}(am_1^{-1})+P_{k-1}(am_2^{-1})\right).\eqnoa\end{equation}
This has the uniform solution as an equilibrium ($k$-independent)
solution. Since in a finite space the uniqueness of equilibrium
is guaranteed, one can show that uniform distribution is indeed approached.
 From this follows the desired claim: Indeed, since $\alpha(q;r)/q$
is identified with the probability that a word fall upon a matrix of
trace $r$ modulo $q$, this probability is clearly given by the
ratio of the number of matrices in $M_q$ having trace $r$ to the total number.

There is still a slight caveat, however: It is well-known that cyclic
behaviour cannot be excluded on general grounds. Indeed, odd-even oscillations
are in fact observed for $q$ equal to two. Nevertheless, it is readily
seen that the effect of such oscillations dies out when one averages over $k$.
Thus, in the absence of oscillations, it would be sufficient to take the
average over one set of conjugacy classes defined by a sufficiently large
value of $k$. In  the presence of oscillations we must still average over
different values of $k$. In fact, it could probably be shown that
cyclic behaviour for cycles larger than two cannot exist. It certainly has
never been observed up to now in the specific cases we have looked
at. One might add that the transition matrix of the above Markov chain
has some very remarkable properties: Its eigenvalues are highly degenerate
and appear to lie on very specific loci of the complex plane. A great deal
of this unexpected structure can be traced back to the fact that this matrix
is invariant under the action of the modulary group (the modular group
taken modulo $q$) and is related to the regular representation of the
latter which is always reducible. We do not go any further
into those details, however, because they
are unnecessary to our immediate purpose.

\appendice{B}

Here we give another derivation of the Hardy--Littlewood
method, which has the advantage of highlighting the
approximations involved.
We need to evaluate the integral on the left-hand side
of eq.~(\ref{hardy}) which expresses the
function $\gamma(r)$. We therefore
divide the unit circle in intervals $I_{p,q}$ centered around
$\exp(2\pi ip/q)$, where $p$ and $q$ run over all relatively prime numbers
with $p<q$ and $q$ less than some prescribed upper bound $Q$ which
later goes to infinity. If one now divides the integral in this way
and denotes the interval $I_{p,q}$ shifted so as to be centered
around $z=1$ by $J_{p,q}$, one obtains:
\begin{eqnarray}
\eqnoa
\gamma(r)&=&\lim_{u\to0}(2u)\sum_{(p,q)=1}\sum_{n,\pr{n}}\alpha(n)
\alpha(\pr{n})e^{-(n+\pr{n})u}\\
&&\qquad \int_{J_{p,q}}{d\psi\over2\pi}
\exp\left(\tpipq(n-\pr{n}-r)-i(n-\pr{n}-r)\psi\right).\nonumber
\end{eqnarray}

To simplify the expression, one rewrites the sums over $n$ and $\pr{n}$
as sums over $\pr{m},\pr{r}$ and $\ppr{m},\ppr{r}$ respectively, where
$n=\pr{m}q+\pr{r}$ and $\pr{r}$ is between 0 and $q-1$,
and similarly for $\pr{n}$.
In this case, the integral over $J_{p,q}$ is strongly oscillatory
for $\pr{m}\neq\ppr{m}$ and one finds:
\begin{eqnarray}\eqnoa
\gamma(r)&=&\lim_{u\to0}{2u\over q}\sum_{(p,q)=1}\sum_{\pr{m}}
\sum_{r,\pr{r}=0}^{q-1}\alpha(\pr{m}q+\pr{r})\alpha(\pr{m}q+\ppr{r})\\
&&\qquad\exp\left(-(2\pr{m}q+\pr{r}+\ppr{r})u\right)e^{\tpipq(\pr{r}
-\ppr{r}-r)}.\nonumber
\end{eqnarray}
At this stage we make a fairly tricky approximation: In essence, we
assume that there are no other correlations between the $\alpha(n)$ than
those implied by the dependence on $q$ and $r$ of its average over
numbers which are equal to $r$ modulo $q$. In that sense, we rewrite
eq.~\last{} as:
\begin{eqnarray} \eqnoa
\gamma(r)&=&q^{-2}\sum_{(p,q)=1}
\sum_{r,\pr{r}=0}^{q-1}\alpha(q;\pr{r})\alpha(q;\ppr{r})
\exp\left(\tpipq(\pr{r}-\ppr{r}-r)\right)\nonumber\\
&=&\sum_{(p,q)=1}q^{-2}\left\vert\sum_{\pr{r}=0}^{q-1}\alpha(q;\pr{r})
\exp\left(\tpipq\pr{r}\right)\right\vert^2\exp\left(-\tpipq r\right).
\end{eqnarray}
Perhaps some examples of sequences $\alpha(n)$ for which the method works, and
others for which it fails may be helpful. For example, let us consider
the sequence $\alpha_p(n)$ which is equal to $1$ when $n\equiv0\pmod{2p}$
and $-1$ when $n\equiv p\pmod{2p}$ where $p$ is a given prime, and zero
otherwise. It is easy to check that in this case $\alpha(q;r)$ is
zero unless $q$ is a multiple of $2p$ and $r$ a multiple of $p$.
The complete calculation of $\gamma(r)$ for this sequence by
the formulae given in the text gives complete agreement with the
exact result. On the other hand, let us now randomize this sequence in
the following way: The numbers $\alpha((2k+1)p)$ and $\alpha(2kp)$
are randomly interchanged
with probability $1/2$. The resulting sequence is clearly still
correlated. In particular, $\gamma(p)$ is non-zero. Nevertheless
it is seen that all $\alpha(q;r)$ of this sequence are zero. Thus the
Hardy--Littlewood method clearly fails to take into account mere
short-range correlations. It is seen that adding this sequence
to another, say the $\alpha(n)$ we have been studying
in the text, clearly modifies
the correlations $\gamma(r)$ but in no way affects the $\alpha(q;r)$,
so that one really must assume that there is no short-range correlations
of this type present. Can this assumption be sustained for the $\alpha(n)$
we have been studying?

At first
sight, this appears very strange, since we are dealing with correlations
over quite short ranges. However, from Appendix A, we see that the natural
variable in which to study the development of correlations is not the trace
$n$ but rather the number of symbols necessary to generate a given conjugacy
class. This number, however, does not vary smoothly at all with trace and the
overwhelming majority of conjugacy classes with nearby traces have quite
different number of symbols that generate them. Thus it is allowable to
consider them as decorrelated as the Hardy--Littlewood method
implicitly does.

\appendice{C}
\setcounter{equation}{0}

In the following we develop some basic tools to compute expressions
for $\mq$ and $\mqr$. The fundamental identity we shall be using
is the following easily verified fact:
\begin{equation}
q^{-1}\sum_{r=0}^{q-1}\exp\left(2\pi irx\over q\right)=\delta_{x,0},
\eqnoa
\end{equation}
where $x$ is an integer between $0$ and $q-1$, or else an arbitrary
integer taken modulo $q$. We therefore find, for $\beta(p,q)$:
\begin{eqnarray}
\beta(p,q)&=&{1\over q}\sum_{r=0}^{q-1}\alpha(q;r)\exp\left(\tpipq r\right)
\nonumber\\
&=&{1\over\mq}\sum_{r=0}^{q-1}\mqr\exp\left(\tpipq r\right)
\eqnoa
\end{eqnarray}
$|M_{q,r}|$ is the number of matrices
$\left(\matrix{a&b\cr c&d\cr}\right)$
where $a,b,c,d$ are taken modulo $q$ such that $ad-bc=1$ and $a+d=r$ modulo
$q$.

Therefore
\begin{eqnarray}
\eqnoa
&\beta (p,q)&={1\over\mq}\sum_{r=0}^{q-1}\sum_{abcd=0}^{q-1}\delta_{ad-bc-1,0}
\delta_{a+d-r,0}\exp\left(\tpipq r\right)\\
&=&{1\over q^2\mq}\sum_{r=0}^{q-1}\sum_{abcd=0}^{q-1}\sum_{m\pr{m}=0}^{q-1}
\exp\left(\tpiq\left[m(ad-bc-1)+\pr{m}(a+d-r)+pr\right]\right).\nonumber
\end{eqnarray}
Note that, in the following, the Kronecker deltas must
{\it always\/} be interpreted as being $1$ if the two indices
are equal modulo $q$. We keep the evaluation of $\mq$ for
the end and perform the sums over $\pr{m}$ and $r$ first.
This gives:
\begin{equation}
\beta(p,q)={1\over\mq}\sum_{bcd=0}^{q-1}\sum_{m=0}^{q-1}\delta_{md+p,0}
\exp\left(\tpiq(dp-mbc-m)\right).\eqnoa\end{equation}
Now we use the fact that $(p,q)=1$ from which follows that the equation
$md\equiv-p\pmod q$ cannot hold unless $(d,q)=(m,q)=1$. Under these
circumstances, on the other hand, the equation has a unique solution,
namely $m\equiv-pd^{-1}\pmod q$, where the inverse is to be taken
modulo $q$. It is well-known that if $(d,q)=1$, this inverse exists
and is unique. From this follows, after performing all sums:
\begin{equation}
\beta(p,q)={q\over\mq}\sum_{(d,q)=1}\exp\left(\tpipq(d+d^{-1})\right).
\eqnoa\end{equation}
Note that such a formula cannot be used to obtain, say, $\mq$
by setting $p$ to zero, since the assumption that $(p,q)=1$ is essential
in deriving eq.~\lasta.

This formula gives an explicit expression of the Fourier transformation of
number of matrices modulo $q$ with determinant equals 1. For billiard problems
we shall need also the same quantity but for matrices with determinant $-1$.
Generalizing the previous arguments we obtain:
\begin{equation}
\beta^{(\eps)}(p,q)={q\over\mq}\sum_{(d,q)=1}
\exp\left(\tpipq(d+\eps d^{-1})\right),
\eqnoa\end{equation}
where $\eps=\pm 1$ corresponds to matrices with determinant  $\pm 1$.

If one introduces the so-called Kloosterman sums \cite{{kloos},{USSR}}
\begin{equation}
\eqnoa
S(n,m,c)=\sum_{(d,c)=1}\exp\left({2\pi i\over c}(nd+md^{-1})\right),
\end{equation}
then $\beta^{(\eps)}(p,q)$ can be written as
\begin{equation}
\eqnoa
\beta^{(\eps)}(p,q)={q\over\mq}S(p,\eps p,q)
\label{kloos}
\end{equation}
It is easy to see that $\beta^{(\eps)}(p,q)\to0$ as $q\to\infty$.
More specifically, It has been shown that \cite {weil}:
\begin{equation}
\eqnoa
|S(n,m,c)|<d(c)\sqrt{c}(m,n,c)^{1/2}
\end{equation}
where $d(c)$ is the number of divisors of $c$ and $(m,n,c)$ is the largest
common divisor of $m,n,q$. Such estimates
yield sharp bounds for the $\beta(p,q)$.

To finish, we need the value of $\mq$. This is found as follows:
First, we can limit ourselves to the case of $q=s^n$, where $s$ is a
prime number, because $\mq$ is a multiplicative function of $q$
by the Chinese Remainder Theorem. Consider first the case $q$
equal to $s$. In this case, we are dealing with matrices over a field.
The number of singular matrices is easily seen to be $s^3+s^2-s$.
Indeed, there are $s^2-1$ different ways of choosing the first row to
be a non-zero vector, and to each of those correspond $s$ different vectors
for the second row, which must be taken proportional to the first. If the
first row is zero, on the other hand, any choice for the second row
will yield a singular matrix. Thus the number of regular matrices
is found to be $(s-1)s(s^2-1)$. These are distributed uniformly
over $s-1$ different non-zero values of the determinant, so that
\begin{equation}
\left|M_s\right|=s(s^2-1)=s^3(1-s^{-2}).\eqnoa\end{equation}
Now consider the case $q=s^n$. Any numbers $a$, $b$, $c$ and $d$
satisfying $ad-bc\equiv1\pmod{s^n}$ will also satisfy the same
equation modulo $s^{n-1}$. Let us then define
\begin{equation}
a\equiv a_1+\alpha s^{n-1}\pmod{s^n},\eqnoa\end{equation}
where $a_1$ is $a$ taken modulo $s^{n-1}$ and $\alpha$ is an appropriate number
taken modulo $s$. Define similar numbers for $b$, $c$ and $d$.
One then has
\begin{equation}
ad-bc=a_1d_1-b_1c_1+s^{n-1}(a_1\delta+d_1\alpha-
b_1\gamma-c_1\beta)\pmod{s^n}.\eqnoa\end{equation}
Hence, to obtain a solution modulo $s^n$, we need an arbitrary
solution modulo $s^{n-1}$ and a set of numbers $\alpha$, $\beta$,
$\gamma$ and $\delta$ satisfying
\begin{equation}
a_1\delta+d_1\alpha-b_1\gamma-c_1\beta\equiv0\pmod s. \eqnoa\end{equation}
But it is impossible that $a_1$, $b_1$, $c_1$ and $d_1$ should
all be zero. Therefore, let us assume, say, that $a_1\ne0$. Then all
possible sets of values for $\alpha$, $\beta$ and $\gamma$ yield
exactly one value for $\delta$. Therefore to each solution modulo
$s^{n-1}$ there correspond exactly $s^3$ solutions modulo
$s^n$, from which follows, for $q=s^n$
\begin{equation}
\mq=s^3\left|M_{q/s}\right|=s^{3n}(1-s^{-2}).\eqnoa\end{equation}
 From this follows the general relation (see e.g. \cite{arith,rankin})
\begin{equation}
\mq=q^3\prod_{p|q}(1-p^{-2}).\eqnoa\end{equation}

\appendice{D}
\setcounter{equation}{0}

In this Appendix, we give an expression for $A_r(q)$ which is defined
by the following relation
\begin{equation}
A_r(q)=\sum_{p:(p,q)=1}\left|\beta(p,q)\right|^2\exp\left(\tpipq r
\right).\eqnoa\end{equation}
Using the above expression for $\beta(p,q)$ one obtains:
\begin{equation}
A_r(q)={q^2\over\mq^2}a_r(q),
\eqnoa
\end{equation}
where
\begin{equation}
a_r(q)=\sum_{p:(p,q)=1}\sum_{(d,q)=1}\sum_{(\delta,q)=1}
\exp\left(\tpipq(d+d^{-1}-\delta-\delta^{-1}-r)\right).\eqnoa\end{equation}
Now we can limit ourselves to evaluating $a_r(q)$ for $q$ equal to $s^n$
and $s$ a prime, since $a_r(q)$ is in fact a multiplicative
function of $q$. Let us first consider the case where $q$ is equal to
$s$ and $s$ is odd. In this case we have:
\begin{eqnarray}
\eqnoa
a_r(s)&=&\sum_{p=1}^{s-1}\sum_{d=1}^{s-1}
\sum_{\delta=1}^{s-1}
\exp\left(\tpips(d+d^{-1}-\delta-\delta^{-1}-r)\right)\nonumber\\
&=&\sum_{p=0}^{s-1}\sum_{d=1}^{s-1}\sum_{\delta=1}^{s-1}
\exp\left(\tpips(d+d^{-1}-\delta-\delta^{-1}-r)\right)\\
&&\qquad-(s-1)^2.\nonumber
\end{eqnarray}
Now one notices that the sum appearing in the last equation
of Eq.~\lasta{} denotes, up to a factor
of $s$, the number of solutions of the equation
\begin{equation}
d+d^{-1}-\delta-\delta^{-1}\equiv r\pmod s.\eqnoa\end{equation}
One sees further that $x$ can be represented in the form $d+d^{-1}$
modulo $s$ if and only if $x^2-4$ can be represented as a square. This
follows from the fact that the equation
\begin{equation}
d+d^{-1}\equiv x\pmod s\eqnoa\end{equation}
has the solution
\begin{equation}
d\equiv{1\over2}\left(x\pm\sqrt{x^2-4}\right)\pmod s.\eqnoa\end{equation}
Here again $1/2$ must be taken modulo $s$, but since $s$ is an odd
prime this is always possible. Further, we see that the number of different
representations of $x$ as $d+d^{-1}$ is $1+((x^2-4)/q)$, where we
are using the Legendre symbol already defined in the text. From this
we obtain the following representation of $a_r(s)$:
\begin{eqnarray} \eqnoa
a_r(s)&=&s\sum_{x=0}^{s-1}\left(1+
\jac{(x^2-4)}{s}\right)\left(1+\jac{((x+r)^2-4)}{s}\right)-(s-1)^2
\nonumber\\
&=&s(s-2)+s\sum_{x=0}^{s-1}
\jac{(x^2-4)((x+r)^2-4)}{s}-(s-1)^2\\
&=&s\sum_{x=0}^{s-1}
\jac{(x^2-4)((x+r)^2-4)}{s}-1\nonumber
\end{eqnarray}
where we have made use of the following standard identity
\begin{equation}
\sum_{x=0}^{s-1}\jac{x(x-a)}{s}=-1,\eqnoa\end{equation}
which we prove for completeness in Appendix E which gives
some elementary properties of the Legendre symbol.
For all practical purposes, this representation is sufficiently
explicit, and no more specific one could be found for general
values of $r$. For $r=0$ and $r=\pm4$, however, closed expressions
can indeed be found: We shall only require the case $r$ equal to
zero. In this case the remaining sum involving Legendre symbols
becomes $s-2$, since these
are identically equal to one except for $x=\pm2$.
 From this we finally obtain for $s$ an odd prime
\begin{equation}
A_0(s)={s^2-2s-1\over s^4(1-s^{-2})^2}.\eqnoa\end{equation}

Now let us consider the case of $q$ equal to $s^n$
In this case, we reexpress eq.~(\appnum.2) in the following way:
Let $B_r(q)$ be the number of roots of the equation
\begin{equation}
d+d^{-1}-\delta-\delta^{-1}\equiv r\pmod q.\eqnoa\end{equation}
In other words:
\begin{equation}
B_r(q)=q^{-1}\sum_{p=0}^{q-1}\sum_{(d,q)=1}\sum_{(\delta,q)=1}
\exp\left(\tpipq(d+d^{-1}-\delta-\delta^{-1}-r)\right).\eqnoa\end{equation}
$a_r(s^n)$ was defined as a sum over $p$ coprime to $s^n$. It is evident that
it is equivalent to the sum over all $p$ minus sum over $p$ divisible on $s$.
 From this follows that we can express $a_r(q)$ in terms
of $B_r(q)$:
\begin{eqnarray} \eqnoa
a_r(s^n)&=&s^nB_r(s^n)-s^2s^{n-1}B_r(s^{n-1})\nonumber\\
&=&s^n\left(B_r(s^n)-sB_r(s^{n-1})\right).
\end{eqnarray}
The factor $s^2$ in the second term comes from the fact
that the summation over $d$ and $\delta$ in $a_r(s^n)$ runs
from $0$ to $s^n-1$, whereas in $B_r(s^{n-1})$ it only runs
from $0$ to $s^{n-1}-1$.

We must therefore evaluate the number of roots of eq.~\backa{2}.
There is an important simplification, however: Since we must eventually
evaluate the difference between the number of roots of eq.~\backa{2}
for $s^n$ and $s$ times the number of those roots for $s^{n-1}$,
we can neglect all roots modulo $s^{n-1}$ which generate exactly $s$ roots
modulo $s^n$, since these then cancel exactly. In the following
we shall always associate to a root modulo $s^n$ the corresponding
number modulo $s^{n-1}$, which is also a root of the equation.

This means that we can immediately discard all roots in which we do
not have both $d\equiv\pm1\pmod s$ and $\delta\equiv\pm1\pmod s$. Indeed,
let either of those two be different from $1$ modulo $s$; for definiteness
let it be $d$. In this case we have
\begin{equation}
d=d_1+\alpha s^{n-1}\qquad\delta=\delta_1+\beta s^{n-1},\eqnoa\end{equation}
where $d_1$ is a number modulo $s^{n-1}$, $\alpha$ is a number
modulo $s$ and similarly for $\delta_1$ and $\beta$. In fact, we
shall systematically use suffixes and Greek letters according to
this convention. Putting this into eq.~\backa{3} and developing the terms
in $s^{n-1}$ as first order infinitesimals (which one may do
since they have the same property modulo $s^n$ of being different
from zero but vanishing in any power higher than the first) one
obtains
\begin{equation}
\alpha(1-d_1^{-2})-\beta(1-\delta_1^{-2})\equiv 0\pmod s. \eqnoa\end{equation}
 From this follows that any value of $\beta$ determines $\alpha$ uniquely
and that $\beta$ can be chosen arbitrarily.
Thus there are exactly $s$ solutions modulo $s^n$ to every such solution
modulo $s^{n-1}$, so that only solutions with
both $d\equiv\pm1\pmod s$ and $\delta\equiv\pm1\pmod s$ need be considered.

In this case we can express $d$ and $\delta$ in the following way
\begin{equation}
d=\pm1+\sum_{k=1}^{n-1}\alpha_k s^k\qquad
\delta=\pm1+\sum_{k=1}^{n-1}\beta_k s^k.\eqnoa\end{equation}
Again, we can put these expressions in the equation to be solved
treating the expression as a formal power series which is cut off
at order $n$. Two cases appear: Either all $\alpha_k$ and $\beta_k$
which would appear in terms higher than linear modulo $s^{n-1}$
vanish or they do not.
In the latter case, a straightforward generalization of the above argument
shows that every solution modulo $s^{n-1}$ generates $s$ solutions
modulo $s^n$, so that these can again be discarded. Finally, there remains
the case in which no terms but the linear ones are non-zero. In this case
all these terms cancel, so that we are led to the conditions
$r\equiv0\pmod{s^{n-1}}$ or $r\equiv\pm4\pmod{s^{n-1}}$ for $a_r(s^n)$
to be different from zero.

We must now distinguish between the case where $n$ is even or odd. Let
us first consider the even case. We then define $n$ to be $2k$ and consider
first the case $r$ equal to zero. This means we can set
\begin{equation}
d=\pm1+\sum_{l=k}^{n-1}\alpha_l s^l\qquad
\delta=\pm1+\sum_{l=k}^{n-1}\beta_l s^l,\eqnoa\end{equation}
where the signs of the leading $\pm1$ must now be taken equal, so that $r$
will be zero. In this case the solutions modulo $s^n$ are simply all
possible choices of $\alpha_l$ and $\beta_l$ over the prescribed range,
since they automatically cancel. That is, there are $2s^n$ solutions,
where the factor $2$ comes from the possibility to choose
both signs for the leading term. Similarly, one obtains for the
same numbers taken modulo $s^{n-1}$ that there are $2s^{n-2}$
solutions, from which follows
\begin{equation}
a_0(s^n)=2s^n(1-s^{-1}).\eqnoa\end{equation}
Similarly, for $n$ even, it is easy to see that
\begin{equation}
a_{\pm4}(s^n)=s^n(1-s^{-1}),\eqnoa\end{equation}
since we have no possibility of choosing two signs. The rest of the
argument is exactly as above, however.

If we now consider the case where $r$ is zero or $\pm4$ modulo
$s^{n-1}$ but not modulo $s^n$, we find the following:
It is not possible to have satisfy the equation (C.10)
modulo $s^n$ with numbers of the form given by eq.~\backa{2}.
This implies that for $t\neq0$:
\begin{equation}
a_{ts^{n-1}}(s^n)=-2s^n s^{-1},
\eqnoa\end{equation}
as well as
\begin{equation}
a_{\pm4+
ts^{n-1}}(s^n)=-s^n s^{-1}.
\eqnoa\end{equation}

Let us now consider $n$ odd, that is, equal to $2k+1$.
In this case we must take into
account all solutions of the form:
\begin{equation}
d=\pm1+\sum_{l=k}^{n-1}\alpha_l s^l\qquad
\delta=\pm1+\sum_{l=k}^{n-1}\beta_l s^l,\eqnoa\end{equation}
If we now compute $d^{-1}$ from this expression, we find
\begin{equation}
d^{-1}=1-\sum_{l=k}^{2k}\alpha_l s^l+\alpha_{k}^2s^{2k}.\eqnoa\end{equation}
Putting the expression in eq.~\lasta{} into the equation, we find that
\begin{equation}
%% FOLLOWING LINE CANNOT BE BROKEN BEFORE 80 CHAR
d+d^{-1}-\delta-\delta^{-1}=(\alpha_{k-1}^2-\beta_{k-1}^2)s^{n-2}.\eqnoa\end{equation}
Therefore, the number of solutions of this equation is the number of
zeroes of the expression $\alpha_k^2-\beta_k^2$, which is $2s-1$
multiplied by the number of possible choices of the remaining parameters.
We therefore have
\begin{equation}
B_0(s^{2k+1})-sB_0(s^{2k})=2(2s-1)s^{2k}-2s^{2k+1}.\eqnoa\end{equation}
 From this one obtains again
\begin{equation}
a_0(s^n)=2s^n(1-s^{-1}),\eqnoa\end{equation}
so that the expression in the case of $r$ equal to zero is unchanged.
In the case of $r$ equal to $\pm4$, we must consider the number
of times the expression $\alpha_k^2+\beta_k^2$ becomes zero. This
either occurs $2s-1$ times or once only, depending on whether
$-1$ can be expressed as a square modulo $s$, which, as is
well-known, depends on whether $s$ is equal to $1$ or $-1$
modulo $4$. Thus, if $s\equiv1\pmod4$ the above formulae are also
recovered in the case of odd $n$, whereas in the case of $s\equiv-1
\pmod4$ one obtains
\begin{equation}
B_{\pm4}(s^{2k+1})-sB_{\pm4}(s^{2k})
=s^{2k}-s^{2k+1},\eqnoa\end{equation}
from which follows that generally speaking
\begin{equation}
a_{\pm4}(s^n)=\eps(n,s)s^n(1-s^{-1}),\eqnoa\end{equation}
where $\eps(n,s)$ is $1$ either if $n$ is even or $s$ of the type
$4m+1$ and $-1$ in the other case.

It now remains to check the formulae in the only remaining case,
which is when $r$ is equal to zero or $\pm4$ modulo $s^{n-1}$
but not modulo $s^n$, still  with $n$ odd. Under these
circumstances the entire reasoning
shown above can be repeated word for word, except that at the point
where we evaluate how often $\alpha_k^2-\beta_k^2$ takes the value zero,
we must now ask how many times it takes on a value different from zero.
As we shall see, this last is independent of the value considered
and is $s-1$. This is verified as follows: Let $N$ be the number
of ways in which $w$ can be expressed as the difference of two squares.
One finds
\begin{eqnarray} \eqnoa
N&=&\sum_{x=0}^{s-1}\left(1+\jac{x}{s}\right)\left(1+\jac{x+w}{s}\right)
\nonumber\\
&=&s-1,
\end{eqnarray}
where we have again used the identity (C.8) to simplify the expression.
 From this we obtain
\begin{equation}
B_{ts^{n-1}}(s^{2k+1})-sB_0(s^{2k})
=2(s-1)s^{2k}-2s^{2k+1}=-2s^{2k},\eqnoa\end{equation}
whereas for $\pm4+ts^{n-1}$ we must ask how often the number
$\alpha_k^2+\beta_k^2$ takes on a fixed non-zero value.
Again, depending on whether $s\equiv1\pmod4$ or not, it takes
the value $s-1$ times or $s+1$ times as is seen by looking at the
sum
\begin{equation}
\sum_{x=0}^{s-1}\left(1+\jac{x}{s}\right)\left(1+\jac{a-x}{s}\right).
\eqnoa
\end{equation}
The remaining case $s=2$ can be treated analogously. The results are presented
in Eqs.~(\ref{2prime})-(\ref{2power}).

The above--discussed values of $A_r(q)$ correspond to matrices with determinant
+1. For billiard problems we need also matrices with determinant $-1$.
The corresponding formulae for $\beta^{(\pm)}(p,q)$ are presented
in Appendix B. Instead of one function $A_r(q)$ one has 3 functions
\begin{equation}
A_r^{(\eps_1,\eps_2)}(q)=\sum_{p:(p,q)=1}\beta(p,q)^{(\eps_1)}
\beta(p,q)^{(\eps_2)}\exp\left(\tpipq r\right).
\eqnoa
\end{equation}
$A_r^{(++)}(q)=A_r(q)$ and the two other functions can be computed in the same
fashion
as $A_r(q)$. We omit the details of the calculations and present only final
results.

When $q=s$ is a prime ($s\neq 2$)
\begin{equation}
A_r^{(\eps_1,\eps_2)}(q)={s^2\over\ms^2}\left[s\sum_{x=0}^{s-1}
\jac{(x^2-4\eps_1)((x+r)^2-4\eps_2)}{s}-1\right].
\eqnoa
\end{equation}
For later applications the value of $A_0^{(\eps_1,\eps_2)}(s)$ is important.
One obtains:
\begin{eqnarray}
\eqnoa
A_0^{(--)}(s)&=&A_0^{(++)}\qquad\hbox{ if } s=4m+1 \nonumber\\
             &=&\frac{1}{s^2}\qquad\hbox{ if } s=4m+3,
\end{eqnarray}
\begin{equation}
A_0^{(+-)}(s)=\frac{1}{s^4(1-s^{-2})}(sH_s(1)-s-1)
\eqnoa
\label{A+-}
\end{equation}
where
$$ H_s(k)=\sum_{x=0}^{s-1}\left (\frac{x(x^2-k)}{s}\right ),$$
$H_s(k)=0$ if $(-1/s)=-1$.

The explicit formula for $H_s(1)$ is not known but it is possible to show
\cite{vino} that if $(-1/s)=1$
$$(H_s(1))^2+(H_s(\alpha))^2=s$$
where $\alpha$ is a number for which $(\alpha/s)=-1$. Therefore for any $s$
$$ |H_s(1)|<\sqrt{s}.$$

For $q=s^n$ $(n\geq 2)$ it follows that if $(-1/s)=-1$ (i.e. there is no
solution
of the equation $\chi^2\equiv -1$   mod $s$) than
$$A_{r}^{(--)}(s^n)=A_{r}^{(+-)}(s^n)=0.$$
If $(-1/s)=+1$
so that there is a number $\chi$ such that $\chi^2\equiv -1$ mod $s$, then
$$A_{r}^{(--)}(s^n)=A_{\chi r}^{(++)}(s^n).$$
And finally, if $r\neq \pm 1 \pm \chi$, then $A_{r}^{(+-)}(s^n)=0$.

An important consequence of these relations is that for all $s$
$$A_{0}^{(+-)}(s^n)=0.$$

\appendice{E}
\setcounter{equation}{0}

In this Appendix we give some standard properties of the Legendre symbol.
Since multiplication modulo $q$ is a group when $q$ is a prime, one has the
well known identity
\begin{equation}
x^{q-1}=1\qquad(x\neq0)\eqnoa
\end{equation}
 From this follows that if $x$ is a square and is non-zero
modulo $q$ one has $x^{(q-1)/2}$ equal to one, whereas for $x$
arbitrary it can only take on the values $1$ and $-1$.
On the other hand,
one finds that the squares of all positive numbers between $1$ and
$(q-1)/2$ are in fact distinct, so that there are at least
$(q-1)/2$ different squares. But a polynomial of degree $n$
(with non-zero leading coefficient) cannot have more than
$n$ different roots. We therefore have the general result
\begin{equation}\eqnoa
\jac{x}{q}=x^{(q-1)/2}\pmod q
\end{equation}
 From this follows
\begin{equation}
\jac{xy}{q}=\jac{x}{q}\jac{y}{q}\eqnoa
\end{equation}
as well as the result that $-1$ has a square root if and only if
$q$ is of the form $4k+1$. Since we have exactly $(q-1)/2$
squares and the same number of non-squares we find the
simple result
\begin{equation}
\sum_{x=0}^{q-1}\jac{x}{q}=0\eqnoa
\end{equation}
Further, one finds, for $a\neq0$:
\begin{eqnarray}
\eqnoa
\sum_{x=0}^{q-1}\jac{x}{q}\jac{x+a}{q}&=&
\sum_{x=1}^{q-1}\jac{x(x+a)x^{-2}}{q}\nonumber\\
&=&\sum_{x=1}^{q-1}\jac{1+ax^{-1}}{q}\\
&=&\sum_{x=0}^{q-1}\jac{x}{q}-1=-1\nonumber\\
\end{eqnarray}
These are the only properties we have used in the text. Much more
is known, however, and can be found in \cite{{vino},{basic}}.

\appendice{F}
\setcounter{equation}{0}

In this Appendix we shall compute the function
\begin{equation}
\eqnoa
G_N(x)=2\pi\!\!\sum_{
\begin{array}{c}
{\scriptstyle(p,q)=1}\\
\scriptstyle0<p/q<x/(2\pi)\\
\scriptstyle q>N
\end{array}
}\!\!\beta(q).
\end{equation}
where $\beta(q)$ is defined as follows: For prime powers $\omega^n$
it has the value $A_0(\omega^n)$, where $A_0(p^n)$ is defined
by Eqs.~(\ref{oddpr}) and (\ref{oddprpower}) in the body of the text.
$\beta(q)$ is then determined everywhere by the multiplicative
property. Let
\begin{equation}
q=\omega_1^{n_1}\omega_2^{n_2}\ldots \omega_k^{n_k}\eqnoa
\label{decomp}
\end{equation}
is a canonical representation of integer $q$ into a product of primes
$\omega_i$ then
\begin{equation}
\beta(q)=\frac{1}{q^3}\prod_{\omega_i}\frac{B(\omega_i ,n_i)}
{(1-\omega_i^{-1})(1-\omega_i^{-2})^2}
\eqnoa
\end{equation}
where if $\omega\neq 2$
$$B(\omega,1)=1-2/\omega-1/\omega^2,\;\;B(\omega,n)=2(1-1/\omega).$$
For $\omega=2$
\begin{eqnarray}
\eqnoa
B(2,1)&=&1/4,\quad B(2,2)=1/2,\quad B(2,3)=0,\nonumber\\
B(2,4)&=&1,\quad B(2,5)=0,\quad B(2,n)=2\quad\hbox{if
$n\geq 6$}.
\end{eqnarray}
To take into account the inequalities $p/q<x/2\pi$ and $q>N$
we find it convenient to use the following identity:
\begin{equation}
\eqnoa
{1\over2\pi i}\int_{\eps-i\infty}^{\eps+i\infty}{u^{-s}\over s}ds=
\left\{
\begin{array}{ll}
0&\hbox{if $u>1$}\\
1&\hbox{if $0<u<1$}\\
\end{array}
\right.
\end{equation}
where $\eps>0$. We therefore introduce the function
\begin{equation}
\eqnoa
\hat G(s,\pr{s})=\left({x\over2\pi}\right)^s\frac{1}{N^{\pr{s}}}
\sum_{(p,q)=1}\frac{q^{s+\pr{s}}\beta(q)}{p^s}.
\end{equation}
One then obtains $(\eps_1,\eps_2>0)$
\begin{equation}
G_N(x)=\frac{1}{(2\pi i)^2}\int_{\eps_1-i\infty}^{\eps_1+i\infty}\frac{ds}{s}
\int_{\eps_2-i\infty}^{\eps_2+i\infty}\frac{d\pr{s}}{\pr{s}}
\hat G(s,\pr{s})
\eqnoa
\label{int}
\end{equation}

The summation in $\hat G(s,\pr{s})$ is taken over
all $q$ and all $p$ coprime. The
latter sum can be computed using the standard formula (see e.g. \cite{hardy})
\begin{equation}
\sum_{(p,q)=1}f(p)=\sum_{k=1}^{\infty}\sum_{\delta |q}f(k\delta)\mu(\delta),
\eqnoa
\end{equation}
where $\mu(\delta)$ is the M\"obius function defined as a multiplicative
function which is zero on all numbers which are divisible by a square
and satisfies
$$\mu(1)=1\qquad\mu(p)=-1$$
for all primes $p$. This gives
$$\sum_{(p,q)=1}\frac{1}{p^s}=\sum_{k=1}^{\infty}\frac{1}{k^s}\sum_{\delta |q}
\frac{\mu(\delta)}{\delta^s}=\zeta(s)\prod_{\omega |q}(1-\omega^{-s}),$$
where $\zeta(s)$ is the Riemann zeta function and the product
extends over all prime factors.

Therefore
\begin{equation}
\hat G(s,\pr{s})=\left (\frac{x}{2\pi}\right )^s\frac{1}{N^{\pr{s}}}
\zeta(s)\sum_{q}\frac{1}{q^{3-s-\pr{s}}}
\prod_i\frac{1-\omega_i^{-s}}{(1-\omega_i^{-1})(1-\omega_i^{-2})^2}
B(\omega_i,n_i),
\eqnoa
\label{gs}
\end{equation}
where $\omega_i$ and $n_i$ are as in Eq.~(\ref{decomp}).
Let us define
\begin{equation}
\frac{C_k(\omega)}{\omega^k}=\sum_{n=1}^{\infty}\frac{1}{\omega^{nk}}
B(\omega,n),
\eqnoa
\end{equation}
where $k=3-s-\pr{s}$. For $\omega\neq 2$ a direct computation gives
\begin{eqnarray}
\eqnoa
\frac{C_k(\omega)}{\omega^k}&=&\frac{B(\omega,1)}{\omega^k}+
\sum_{n=2}^{\infty}\frac{1}{\omega^{nk}}B(\omega,n)\nonumber\\
&=&\frac{1}{\omega^k}\left(
1-\frac{2}{\omega}-\frac{1}{\omega^2}\right)+
2\left(1-\frac{1}{\omega}\right)
\sum_{n=2}^{\infty}\frac{1}{\omega^{nk}}\\
&=&\frac{1}{\omega^k}\left(1-\frac{1}{\omega^2}-
2\frac{\omega^{k-1}-1}{\omega^{k}-1}\right ).\nonumber
\end{eqnarray}
Therefore if $\omega\neq 2$
\begin{equation}
C_k(\omega)=1-\frac{1}{\omega^2}-2\frac{\omega^{k-1}-1}{\omega^{k}-1}.
\eqnoa
\label{ck}
\end{equation}
Similarly
\begin{equation}
%% FOLLOWING LINE CANNOT BE BROKEN BEFORE 80 CHAR
C_k(2)=\frac{1}{4}+\frac{1}{2^{k+1}}+\frac{1}{2^{3k}}+\frac{1}{2^{4k-1}(2^k-1)}.
\eqnoa
\label{c2}
\end{equation}
We now rewrite Eq.~(\ref{gs}) as a sum over all $\omega_i$ and $n_i$.
One can then perform the sums over the $n_i$ and obtains a factor
$C_k(\omega_i)$ for each $\omega_i$. Finally one obtains
\begin{equation}
\hat G(s,\pr{s})=\left (\frac{x}{2\pi}\right )^s\frac{1}{N^{\pr{s}}}\zeta(s)
\prod_{\omega} \left(1+\frac{1}{\omega^{3-s-\pr{s}}}
\frac{1-\omega^{-s}}{(1-\omega^{-1})(1-\omega^{-2})^2}C_{3-s-\pr{s}}(\omega)
\right).
\eqnoa
\label{Gs}
\end{equation}
where the product is taken over all primes.

This product converges when $1<\Re s<2-\Re \pr{s}$.
To obtain the function $G(x)$ one
should compute the integral (\ref{int}) along the line parallel to the
imaginary axis real part of which lies in the above interval. Because
$N\gg 1$ it is possible to shift the contour of integration over $\pr{s}$
right until it
reaches the first singularity of $\hat G(s,\pr{s})$. It is easy
to see that this
singularity is a pole at $\pr{s}=2-s$ coming from the product over all primes.
Putting $\pr{s}=2-s-\eps$ and assuming that $\eps \rightarrow 0$ one has
\begin{equation}
\hat G(s,\pr{s})=\left (\frac{x}{2\pi}\right )^s\zeta(s)
\prod_{\omega}\left (1+\frac{1}{\omega^{1+\eps}}
\frac{1-\omega^{-s}}{(1-\omega^{-1})(1-\omega^{-2})}C_{1}(\omega)\right ).
\eqnoa
\end{equation}
But from Eqs.~(\ref{ck}) and (\ref{c2}) it follows that
\begin{equation}
C_1(\omega)=1-\omega^{-2}
\eqnoa
\label{c1}
\end{equation}
for all $\omega$ including $\omega=2$. Therefore
\begin{equation}
\hat G(s,2-s-\eps)=\left (\frac{x}{2\pi}\right )^s\frac{1}{N^{2-s}}\zeta(s)
\prod_{\omega}\left (1+\frac{D(\omega)}{\omega^{1+\eps}}\right ).
\eqnoa
\end{equation}
where
$$D(\omega)=\frac{1-\omega^{-s}}{(1-\omega^{-1})(1-\omega^{-2})}$$
and as $\eps \rightarrow 0$
\begin{equation}
\hat G(s,2-s-\eps)\rightarrow \left (\frac{xN}{2\pi}\right )^sN^{-2}\zeta(s)
\zeta(1+\eps)K_s.
\eqnoa
\end{equation}
where
$$K_s=\prod_{\omega}\left(1+\frac{D(\omega)}{\omega}\right)
\left(1-\frac{1}{\omega}\right)$$
because
$$\zeta(s)=\prod_{\omega}(1-\omega^{-s})^{-1}.$$
It is well known that the Riemann zeta function $\zeta (s)$ has a pole at
$s=1$ with unit residue (see e.g. \cite{tit}) and consequently as
$\pr{s}\rightarrow 2-s$
\begin{equation}
\hat G(s,\pr{s})\rightarrow N^{-2}\frac{(xN)^s}{(2\pi)^s}\zeta(s)
\frac{K_s}{2-s-\pr{s}}
\eqnoa
\end{equation}
Integrating over $\pr{s}$ one concludes that in the leading order of $1/N$
\begin{equation}
G_N(x)=\frac{N^{-2}}{2\pi i}
\int_{\eps-i\infty}^{\eps+i\infty}\frac{ds}{s(2-s)}
\zeta(s)\frac{(xN)^s}{(2\pi)^s}K_s.
\eqnoa
\end{equation}
For convergence it is necessary that $1<\eps<2$. Rewriting $(xN)^s$ as
$\exp(-s\ln (xN))$ one easily concludes that if $xN\ll 1$ one can shift the
contour of integration right up to the pole at $s=2$ and
\begin{equation}
G_N(x)\rightarrow \frac{x^2}{48}
\eqnoa
\end{equation}
because $\zeta(2)=\pi^2/6$ and $K_2=1$.
If $xN\gg 1$ one can move the contour of integration only left up to the pole
$s=1$ coming from $\zeta (s)$ and
\begin{equation}
G_N(x)\rightarrow \frac{x}{N}K_1
\eqnoa
\end{equation}
where
$$K_1=\prod_{\omega}(1-\frac{1}{\omega})(1+\frac{1}{\omega(1-\omega^{-2})})=
\prod_{\omega}(1-\frac{1}{\omega(\omega+1)}),$$
and the product is taken over all primes.

In Fig.5 we present the plot of $g(x)$. On a double logarithmic scale,
the $x^2$ behaviour for $x\ll1$ is in evidence. The prefactor
is also found to be correct, as shown by the theoretical line.
As indicated in the figure captions, we then show $g(x)$ in which
the peaks with denominators less than $N$ have been subtracted. In these
one sees an approximately linear behaviour over a broad range of $x$, which
is in complete agreement with the behaviour predicted in Sec.~5
and in this Appendix.

For billiard problems one has to compute functions $G(x)$ with
$\beta^{(\eps_1,\eps_2)}(q)$ defined in Eq.~(\ref{beta}).
Using the values of $A_0^{(\eps_1,\eps_2)}(q)$ presented in Appendix D one
concludes that
$$B^{(--)}(\omega,n)=B^{(++)}(\omega,n)$$
if $(-1/\omega)=1$ and consequently
$$C^{(--)}_k(\omega)=C^{(++)}_k(\omega)$$
for such primes in particular (see Eq.~(\ref{c1}))
$$C^{(--)}_1(\omega)=1-\omega^{-2}.$$
For primes with $(-1/\omega)=-1$
$$B^{(--)}(\omega,1)=1-\omega^{-2},\qquad B^{(--)}(\omega,n)=0
\quad\hbox{ if
$n\geq 2$},$$
It means that for all primes the value of $C^{(--)}_1(\omega)$
is the same as for
$C^{(++)}_1(\omega)$ and the asymptotics of the function $G^{(--)}(x)$
coincides with that of $G^{(++)}(x)$.

In particular as $x\ll 1/N$
\begin{equation}
G^{(--)}(x)\rightarrow \frac{x^2}{48}
\eqnoa
\end{equation}
and when $x\gg 1/N$
\begin{equation}
G^{(--)}(x)\rightarrow \frac{x}{N}K_1
\eqnoa
\end{equation}
For the function $G^{(+-)}(x)$ one obtains from Eqs.~(\ref{A+-})
$$B^{(+-)}(\omega,1)=\frac{1}{\omega^2}(sH_s(1)-s-1)$$
and $B^{(+-)}(\omega,n)=0$ for $n\geq 2$. Therefore
$$C^{(+-)}_k(\omega)=\frac{1}{\omega^2}(sH_s(1)-s-1).$$
But we have mentioned that
$$ |H_s(1)|<\sqrt{s}.$$
It means that the product (\ref{Gs}) will have a
singularity only when $s=2.5$ and as $x\ll 1/N$
\begin{equation}
G_N^{(+-)}(x)={\cal O}(x^{2.5}).
\eqnoa
\end{equation}

\pagebreak

\pagebreak

\begin{center}
{\bf Figure captions}
\end{center}

\begin{description}

\item[Fig. 1] The fundamental domain of the modular group. $T$ denotes the
translation $z\rightarrow z+1$, $S$ is the inversion $z\rightarrow -1/z$
and the arrows connect the boundaries identified under these transformations.
The shaded region is the fundamental domain of the modular billiard.

\item[Fig. 2] Two--point correlation function of the multiplicities of the
periodic orbits of the modular group computed from the knowledge
of whose corresponding matrices have traces up to 8000.

\item[Fig. 3] a) The Fourier transform of the two--point correlation function
of the multiplicities of the periodic orbits for the modular domain.

b) Its integral.

\item[Fig. 4] The schematic picture of the function $\avg{f_N(x)}$ for
$0<x<2\pi$.

\item[Fig. 5] The difference between the exact $G(x)$ and the approximate
formula (\ref{as}) for different values of $N$ which are indicated near the
curves. The middle line corresponds to $N=50$.

\item[Fig. 6] Behaviour of the integral of the two-point correlation formfactor
at small $x$. The dotted line corresponds to the sum of all terms up to $N=500$
and the solid one to $N=1000$. The straight line indicates asymptotics
(\ref{5.25}).

\end{description}


\begin{thebibliography}{99}
\def\jpa#1#2#3{J. Phys. A: Math. Gen. {\bf{#1}}, #2 (19#3)}
\def\jpl#1#2#3{Jour. Physique Lettres {\bf{#1}}, #2 (19#3)}
\def\jmp#1#2#3{J. Math. Phys. {\bf{#1}}, #2 (19#3)}
\def\prl#1#2#3{Phys. Rev. Lett. {\bf{#1}}, #2 (19#3)}
\def\pra#1#2#3{Phys. Rev. A{\bf{#1}}, #2 (19#3)}
\def\prb#1#2#3{Phys. Rev. B{\bf{#1}}, #2 (19#3)}
\def\prc#1#2#3{Phys. Rev. C{\bf{#1}}, #2 (19#3)}
\def\prd#1#2#3{Phys. Rev. D{\bf{#1}}, #2 (19#3)}

\bibitem{hopf} E. Hopf, {\em Ergodentheorie}, Springer-Verlag (1937)
\bibitem{enc} {\em Encyclopedia of Mathematical Sciences}, Vol. 1
Dynamical Systems, eds. D.V. Arnold and V.I. Anosov, Springer Verlag (1988)
\bibitem{balvor}N.L. Balazs and A. Voros, Phys. Rep. {\bf143} 109 (1986)
\bibitem{houches} {\em Proceedings of the Les Houches Summer School}
(1989) Chaos and Quantum Physics Eds. M.J. Giannoni, A. Voros, and J.
Zinn--Justin, North Holand, Amsterdam, 1991
\bibitem{gutzbook} M.C. Gutzwiller,
{\em Chaos in Classical and Quantum Mechanics}
Springer Verlag, New-York (1990)
\bibitem{selberg} A. Selberg, J. Ind. Math. Soc. {\bf20} 47 (1956)
\bibitem{hejhal}D. Hejhal,
{\em The Selberg Trace Formula for $PSL(2,R)$}, Vol. 1
Lecture Notes in Mathematics {\bf548} (1979); Vol. 2, ibid. {\bf1001}
(1983)
\bibitem{gutz} M.C. Gutzwiller, \jmp{12}{343}{71}
\bibitem{bgs} O. Bohigas, M.-J. Giannoni and C. Schmit, \prl{52}{1}{84};
\jpl{45}{L1015}{84}
\bibitem{mehta}M.L. Mehta,
{\em Random Matrices and the Statistical Theory of Energy
Levels}, Academic Press, New-York (1967)
\bibitem{charles}C. Schmit in \cite{houches}
\bibitem{charles1} C. Schmit,
{\em Triangular Billiards on the Hyperbolic Plane:
Spectral Properties}, Report IPNO/TH (1991) (unpublished)
\bibitem{aurich1}R. Aurich and F. Steiner, Physica {\bf D 39}, 169 (1989)
\bibitem{aurich2}R. Aurich and F. Steiner, Physica {\bf D 43}, 155 (1990)
\bibitem{berrytab}M.V. Berry and M. Tabor, Proc. R. Soc. London
{\bf A 356}, 375 (1977)
\bibitem{georgeot} E. Bogomolny, B. Georgeot, M.-J. Giannoni and C. Schmit,
\prl{69}{1477}{92}
\bibitem{steiner}J. Bolte, G. Steil and F. Steiner, \prl{69}{2188}{92}
\bibitem{georgeot2} E. Bogomolny, B. Georgeot, M.-J. Giannoni and C. Schmit,
{\em Arithmetical Chaos}, Preprint IPNO/TH 93-51
\bibitem{bolte}J. Bolte, Ph.D. Thesis DESY (1993)
\bibitem{gelfand} I.M. Gelfand, M.I. Graev and I.I. Pyatetskiii--Shapiro,
{\em Representation Theory and Automorphic Functions}, W.B. Saunders Company,
Philadelphia, London, Toronto (1969)
\bibitem{arith}S. Katok, {\em Fuchsian Groups}, Chicago Lecture in
Mathematics, 1992
\bibitem{berry}M.V. Berry, Proc. R. Soc. London {\bf A400}, 229 (1985)
\bibitem{hardy}G.H. Hardy and J.E. Littlewood, Acta Mathematica
{\bf44}, 1 (1922)
\bibitem{magnus}W. Magnus, {\em Non-Euclidean Tesselations and their Groups},
Academic Press (1974)
\bibitem{harm}A. Terras, {\em Harmonic Analysis on Symmetric Spaces
and Applications}, Springer Verlag, Berlin (1979)
\bibitem{sarnak}P. Sarnak, Journal of Number Theory, {\bf 15}, 229 (1982)
\bibitem{matth}C. Matthies and F. Steiner, \pra{44}{R7877}{91}
\bibitem{huber}H. Huber, Math. Ann. {\bf138}, 1 (1959)
\bibitem{berry2}M.V. Berry in \cite{houches}
\bibitem{bogom}E. Bogomolny and C. Schmit, Nonlinearity {\bf6}, 523 (1993)
\bibitem{aurich}R. Aurich and M. Sieber, {\em An Exponentially Increasing
Spectral Form Factor $K(\tau)$ for a Class of Strongly Chaotic Systems}, DESY
Preprint, DESY 92--171 (1992)
\bibitem{tauber}J. Karamata, J. reine und angewandte Mathematik {\bf164},
27 (1931)
\bibitem{feller} W. Feller, {\em An Introduction to Probability
Theory and its Applications}, vol.~1, New-York John Wiley
and sons (1957); vol.~2 (1971).
\bibitem{vino}I.M. Vinogradov, {\em An Introduction to the Theory
of Numbers}, Pergamon Press; Oxford, London New-York, Paris (1961)
\bibitem{basic}G.H. Hardy and E.M. Wright, {\em An Introduction to
the Theory of Numbers}, fourth edition, Oxford Clarendon
Press (1959)
\bibitem{kloos}H.D. Kloosterman, Acta Math. {\bf49}, 407 (1926)
\bibitem{weil}A. Weil, Proc. Nat. Acad. Sci. USA {\bf34}, 204
(1948)
\bibitem{USSR}N.V. Kuznetsov, Journal of Soviet Mathematics {\bf29}, 1131
(1985)
\bibitem{venkov}A.B. Venkov, {\em Spectral Theory of Automorphic Functions},
Proc. Steklov Institute of Math. {\bf 4} (1982)
\bibitem{bogl}E. Bogomolny and P. Leboeuf, {\em Statistical Properties of the
Zeros of Zeta Functions---Beyond the Riemann Case}, Preprint
{\bf IPNO/TH 93-44}, Nonlinearity 1994, to be published
\bibitem{tit} E.C. Titchmarsh, {\em The Theory of the Riemann Zeta-Function},
Oxford, Clarendon Press, 1951
\bibitem{rankin} R.A. Rankin {\em Modular Forms and Functions}, Cambridge
University Press, Cambridge (1977)
\end{thebibliography}
\end{document}